\documentclass{aa}
\usepackage{graphicx}
\usepackage{txfonts}
\usepackage{color}
\usepackage{xspace}
\usepackage{siunitx}
\usepackage{amsmath}
\usepackage{amssymb}
\usepackage{xcolor}
\usepackage{dashbox}
\usepackage{framed}
\usepackage{lipsum}
\usepackage{placeins}
\usepackage{stfloats}

\usepackage{hyperref}

\hypersetup{
        colorlinks=true, % color of internal links
        breaklinks=true,
        linkcolor=blue, % color of internal links
        citecolor=blue, % color of links to bibliography
        filecolor=blue, % color of file links
        urlcolor=blue,
        unicode=false, % non-Latin characters in Acrobat bookmarks
        pdftoolbar=true, % show Acrobat toolbar?
        pdfmenubar=true, % show Acrobat menu?
        pdffitwindow=false, % window fit to page when opened
        pdfstartview={Fit}, % fits the width of the page to the window
        pdftitle={Extended stellar systems in the solar neighborhood V}, % title
        pdfauthor={Stefan Meingast}, % author
        pdfsubject={Cluster coronae},
        pdfcreator={Stefan Meingast}, % creator of the document
        pdfkeywords={},
        pdfnewwindow=true, % links in new window
        pdfdisplaydoctitle=true % display document title instead of file name
}

\makeatletter
\renewcommand*\aa@pageof{, page \thepage{} of \pageref*{LastPage}}
\makeatother

\newcommand{\code}{\texttt}
\newcommand{\gaia}{\textit{Gaia}\xspace}
\newcommand{\aper}{$\alpha\,\mathrm{Per}$\xspace}

\begin{document}

% ============================================================================ %
% Cite alias for Paper I
\defcitealias{ESSI}{Paper~I}
\defcitealias{ESSII}{Paper~II}
\defcitealias{ESSIII}{Paper~III}
\defcitealias{ESSIV}{Paper~IV}

% ============================================================================ %
\title{Extended stellar systems in the solar neighborhood}
\subtitle{V. Discovery of coronae of nearby star clusters\thanks{The cluster membership table is made available online via CDS.}}

% Discovery of open star cluster coronae
% Coronae of nearby young star clusters
% Discovery of coronae of nearby star clusters

% ============================================================================ %
\author{Stefan Meingast\inst{1} \and
        João Alves\inst{1,2} \and
        Alena Rottensteiner\inst{1}}

\institute{Department of Astrophysics, University of Vienna, T\"urkenschanzstrasse 17, 1180 Wien, Austria 
\\ \email{stefan.meingast@univie.ac.at}
\and
Data Science @ University of Vienna, Währinger Straße 29, A-1090 Vienna
}

% ============================================================================ %
\date{Received June 8, 2020 / accepted September 17, 2020}

% ============================================================================ %
\abstract{We present a novel view on the morphology and dynamical state of ten prominent, nearby ($\leq$ \SI{500}{pc}), and young (${\sim}30-$\SI{300}{Myr}) open star clusters with Gaia DR2: \aper, Blanco~1, IC~2602, IC~2391, Messier~39, NGC~2451A, NGC~2516, NGC~2547, Platais~9, and the Pleiades. We introduce a pioneering member-identification method that is informed by cluster bulk velocities and deconvolves the spatial distribution with a mixture of Gaussians. Our approach enables inferring the true spatial distribution of the clusters by effectively filtering field star contaminants while at the same time mitigating the effect of positional errors along the line of sight. This first application of the method reveals vast stellar coronae that extend for $\gtrsim\,$\SI{100}{pc} and surround the cluster cores, which are comparatively tiny and compact. The coronae and cores form intertwined, coeval, and comoving extended cluster populations, each encompassing tens of thousands of cubic parsec and stretching across tens of degrees on the sky. Our analysis shows that the coronae are gravitationally unbound but largely comprise the bulk of the stellar mass of the populations. Most systems are in a highly dynamic state, showing evidence of expansion and sometimes simultaneous contraction along different spatial axes. The velocity field of the extended populations for the cluster cores appears asymmetric but is aligned along a spatial axis unique to each cluster. The overall spatial distribution and the kinematic signature of the populations are largely consistent with the differential rotation pattern of the Milky Way. This finding underlines the important role of global Galactic dynamics in the fate of stellar systems. Our results highlight the complexity of the Milky Way's open cluster population and call for a new perspective on the characterization and dynamical state of open clusters.}

% ============================================================================ %

% ============================================================================ %
\maketitle

% ============================================================================ %
% ============================================================================ %
\section{Introduction}
\label{sec:introduction}

\begin{figure*}
        \centering
        \resizebox{1.0\hsize}{!}{\includegraphics[]{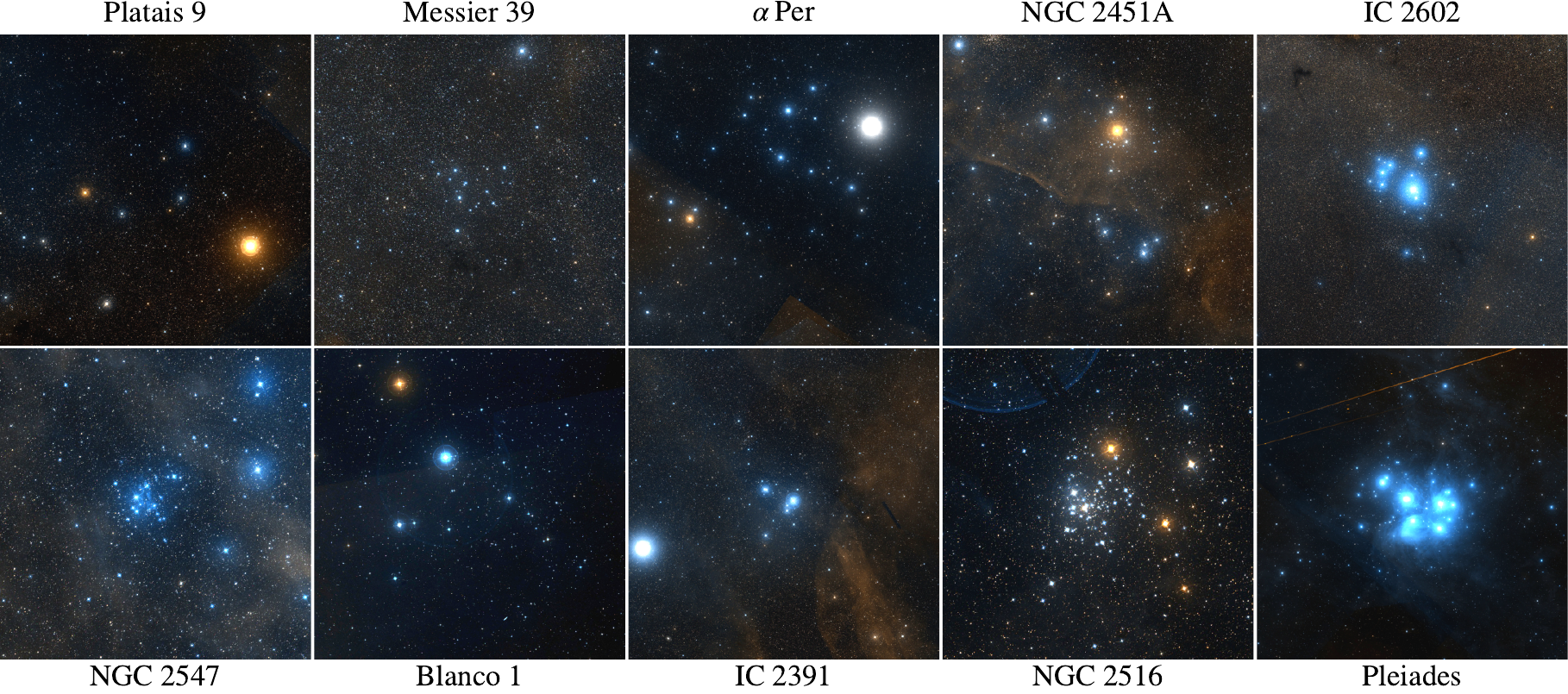}}
        \caption[]{Postage stamps for our ten target clusters, extracted from the Digitized Sky Survey. Each stamp as a side length of \SI{10}{pc} at the distance of the cluster core, and is displayed so that Galactic north is up, and east is left. From top left to bottom right, the clusters are sorted by their peak volume density.}
    \label{img:dss}
\end{figure*}

Star clusters are fundamental building blocks of galaxies. To this day, they are oftentimes thought to be the birthplace of most stars \citep[e.g.,][]{Lada03,Parker07}. Over the past decade, this view has been challenged on multiple occasions (e.g., \citealp{Elmegreen08,Bressert10,Kruijssen12b}; cf. \citealp{Parker12}), where the \gaia mission \citep{gaia_a} and particularly the second data release \citep[\gaia DR2;][]{gaia_dr2} have recently initiated a renaissance in our understanding of the predominant conditions for star formation \citep[e.g.,][]{Wright18,Cantat-Gaudin19b,Kuhn19,Wright19,Ward20}. These insights have been promoted mostly by investigating the large-scale structure and kinematics of massive OB associations with ages up to a few million years. However, exploring the full population of more evolved stellar structures becomes increasingly difficult as stellar systems quickly disperse and blend in with the Galactic field \citep{Lada03,Pellerin07}. As a consequence, the dissociation of young stellar structures presents a challenging obstacle for our understanding of star formation demographics, and in particular, the dynamical evolution of stellar systems.

The dissolution process of open star clusters set inside a Galactic tidal field has been a popular subject in the astrophysical literature in the past decades. The long-term evolution of star clusters has been shown to be dependent on a great variety of variables that include, for instance, galaxy properties, the position in the host galaxy, orbital characteristics, nonaxisymmetric perturbations, tidal stripping, or radial migration \citep[e.g.,][]{Spitzer58,Krumholz19}. On timescales of several hundred million years, dense star clusters dissolve by gradually releasing stars into tidal tails \citep[e.g.,][]{Ernst11,Renaud11}, features that have only recently been confirmed from an observational point of view \citep{ESSI, Roeser19}.

At earlier stages, theory suggests that a crucial phase is the transition of a cluster from an embedded to a gas-free state, usually referred to as residual gas expulsion (e.g., \citealp{Baumgardt07}; cf. \citealp{Dale15}). Briefly summarized, star clusters are not expected to be in virial equilibrium after gas expulsion \citep{Ernst15}. As a consequence, clusters experience a phase of violent relaxation so that individual young stars can become gravitationally unbound and start to expand. Theory establishes the critical factors for the degree of expansion and the disruption of young clusters to be the star formation efficiency \citep[SFE, e.g.,][]{Lada84,Shukirgaliyev17}, the timescales of gas expulsion \citep{Dinnbier20}, the dynamical properties of the cluster at the onset of gas expulsion \citep{Goodwin09}, primordial mass segregation \citep{Brinkmann17}, and subsequent violent relaxation \citep{Shukirgaliyev18}. The relative importance of these factors continues to be openly debated and remains unresolved, mostly because we lack unbiased observational constraints.

Understanding the early evolution of stellar systems is further complicated by the highly substructured nature of star formation. Several studies of young star-forming regions do indeed portray an irregular picture with substantial substructure \citep[e.g.,][]{Cartwright04,Gutermuth08,Kuhn14}. For instance, the nearest star cluster containing massive stars, the Orion nebula cluster, is in fact part of a much larger, highly elongated, substructured, and actively star-forming molecular cloud complex spanning about \SI{90}{pc} \citep{josefa18, Hacar18}. The Orion cloud complex itself is embedded in the recently discovered Radcliffe Wave, an undulating 3 kpc long string of star-forming regions \citep{Alves2020-um}. Particularly in dense environments, \citet{Kruijssen12} suggested that the cluster dissolution process could be dominated by tidal interactions in the greater birth environment of the natal molecular cloud complex.

In particular, a recent discovery highlights the complex interplay between the dynamical evolution of stellar systems and the physical conditions at birth. \citet[][hereinafter referred to as Paper~I]{ESSII} reported the discovery of the large stellar stream Meingast~1 \citep[see also][hereinafter referred to as Paper~IV]{ESSIV} with an age of only about \SI{120}{Myr} \citep{Curtis19}. The stream seems to be more massive than but coeval with the Pleiades star cluster, but appears in stark contrast morphologically: Meingast~1 lacks a prominent cluster core, which means that if it formed as a monolithic cluster, it did not retain a centrally peaked structure. Instead, the population stretches across several hundred parsec and was only recently discovered because homogeneous radial velocity data across the sky became available. As a consequence, Meingast~1 is a prime example that highlights how different physical conditions at early stages can lead to vastly different dynamical evolution scenarios \citep[see also][]{Roeser20}.

Collecting observational evidence for the dissolution process of star clusters has been severely hampered by the inherent difficulties connected to the identification of stripped cluster members (e.g., \citealp{Boss1908,Artyukhina64,Eggen87,Dalessandro15,Yeh19}). The reason for this is that stars located beyond the tidal radius of a system are immersed in an overwhelmingly large number of unrelated field stars. Tracing population members out to several tidal radii and even beyond entails recovering individual sources orders of magnitude below the typical field star density. As a result, observers are confronted with the infamous problem of finding a needle in a haystack. Stripped cluster members cannot be identified with the classical approach of locating overdensities with respect to the background population. Instead, kinematic and chemical information must be used to identify stellar siblings reliably \citep{Kamdar19}. While elemental abundances are only now becoming available in moderately sufficient quantities, the \gaia mission has ushered kinematic measurements into the age of big data. Nevertheless, identifying stellar populations across large regions of the sky remains challenging because \gaia primarily provides proper motion measurements, whereas 3D velocities are available only for a relatively small subset. In particular, projection effects and significant measurement errors in geometric distances still pose delicate challenges even for sophisticated machine-learning applications. In this paper we address these issues and introduce new methods for mitigating common limitations present in popular member-identification and cluster-analysis procedures.

The work presented here focuses on the analysis of ten nearby young open star clusters\footnote{Throughout this paper, the term cluster will be used interchangeably with other expressions (e.g., system, structure) to refer to an entire comoving and coeval population, regardless of the gravitational boundedness.}: \aper (\mbox{\object{Melotte 20}}), \mbox{\object{Blanco 1}}, \mbox{\object{IC 2602}}, \mbox{\object{IC 2391}}, \mbox{\object{Messier 39}}, \mbox{\object{NGC 2451A}}, \mbox{\object{NGC 2516}}, \mbox{\object{NGC 2547}}, \mbox{\object{Platais 9}}, and the \object{Pleiades}. We selected these particular clusters because they are believed to be well known and demonstrate the capabilities of our methods extraordinarily well. Moreover, these clusters span a suitable range in age from about 30 to \SI{300}{Myr} (see Table~\ref{tab:clusters}) so that their dissolution process may be analyzed, and they are located in the solar neighborhood. Figure~\ref{img:dss} shows postage stamps for each cluster from the Digitized Sky Survey\footnote{\href{http://archive.stsci.edu/cgi-bin/dss_form}{http://archive.stsci.edu/cgi-bin/dss\_form}} scaled to an edge length of \SI{10}{pc} at the distance of the cluster core. These clusters all share a prominent appearance on the sky, where some of them have been known for centuries or even millennia. Despite their long-lived history in astrophysical research, we present unambiguous evidence that these apparent clusters comprise only the proverbial tip of the iceberg and are in fact each surrounded by a vast stellar halo, which we refer to as their corona\footnote{The term corona has been used several times throughout the past decades in reference to star clusters, loosely referring to the less dense outer regions of open clusters \citep{Artyukhina64}. Here, we use the term corona to generally describe the stellar population of star clusters beyond their tidal radius.}.

% ============================================================================ %
% ============================================================================ %
\section{Data}
\label{sec:data}

\begin{table}[!t]
\caption{Cluster parameters obtained from the literature.}
\begin{tabular*}{\linewidth}{@{\extracolsep{\fill}} lcccc}
\hline\hline
Name    &       Age     &       $v_R$   &       $v_\phi$        &       $v_Z$ \\
        &       (Myr)   &       (\si{\km \per \second}) &       (\si{\km \per \second})  &       (\si{\km \per \second}) \\
\hline
Platais 9       &       100 (78-347)    &       5.97    &       229.21  &       1.66 \\
Messier 39      &       310 (279-1023)  &       -31.00  &       242.08  &       -5.19 \\
\aper   &       87 (35-110)     &       3.11    &       221.29  &       1.17 \\
NGC 2451A       &       44 (32-148)     &       9.03    &       232.3   &       -4.97 \\
IC 2602 &       35 (30-100)     &       -8.39   &       224.84  &       7.08 \\
NGC 2547        &       27 (27-78)      &       -8.07   &       235.11  &       -3.78 \\
Blanco 1        &       94 (63-209)     &       6.11    &       239.1   &       -2.41 \\
IC 2391 &       36 (26-81)      &       6.82    &       231.27  &       2.09 \\
NGC 2516        &       251 (63-299)    &       -1.34   &       221.13  &       3.06 \\
Pleiades        &       86 (86-176)     &       -5.30   &       217.13  &       -6.52 \\
\hline
\end{tabular*}
\tablefoot{We adopted cluster ages as published by \citet{Bossini19}. The values quoted in parentheses refer to the extrema in the age range as found in the literature. A comprehensive overview of these data is given in Table~\ref{tab:ages} in the appendix. The velocities refer to a galactocentric cylindrical coordinate frame and serve as input for our member-identification method.}
\label{tab:clusters}
\end{table}

We based our analysis exclusively on Gaia DR2. To mitigate the effect of inaccurate measurements and outliers, we used the same quality cuts as employed in \citetalias{ESSIV}. Specifically, the constraints we used are $\varpi / \sigma_\varpi$ > 10, $f_{\mathrm{BP}} / \sigma_{f_{\mathrm{BP}}}$ > 10, $f_{\mathrm{RP}} / \sigma_{f_{\mathrm{RP}}}$ > 10, astrometric\_sigma5d\_max < \SI{0.5}{mas}, visibility\_periods\_used > 6, and the renormalized unit weight error, RUWE < 1.4 \citep[   ][]{ruwe}. Here, $\varpi$, $f_{\mathrm{BP}}$, and $f_{\mathrm{RP}}$ denote the parallax and the measured flux in the BP and RP bands, respectively, while $\sigma_\varpi$, $\sigma_{f_{\mathrm{BP}}}$, and $\sigma_{f_{\mathrm{RP}}}$ refer to the standard error in the corresponding parameters. To obtain a thorough census of the population of each cluster, we further limited parallaxes to values greater than \SI{1}{mas}, corresponding to a geometric distance limit of about \SI{1}{kpc}. These cuts together result in a remarkably clean if conservative sample comprising a total of \num{31682087}{} sources. All further steps in the data analysis, the results, and their analysis are based on this set of sources.

We preferred to calculate distances by inverting the individual parallax measurements as opposed to methods that take various distance biases into account. The reason for this choice was mainly our rigorous constraint of a 10 $\sigma$ significance of the parallax measurements. Furthermore, methods such as described by \citet{bailerjones15}, for example, impose a characteristic length-scale depending on the global galactic field distribution that is not directly applicable to cluster populations. Moreover, we did not apply a systematic parallax offset either as detailed in \citet{gaia_dr2_astrometric_solution}, for example, because the quoted value is a global average and can be significantly different on local scales.

In the main text of this paper, we refer to a series of parameters and reference frames. Most notably, $XYZ$ coordinates and their time differentials $UVW$ refer to Cartesian coordinates in a Galactic reference frame centered on the Sun, where the velocity components are not corrected for the solar motion. In this frame, $X$ and its differential $U$ point toward the Galactic center, $Y$ and $V$ point in the direction of Galactic rotation, and $Z$ and $W$ are positive toward the Galactic north pole. In some cases, we prefer Galactocentric cylindrical coordinates with $v_R$, $v_\phi$, and $v_z$ that denote the radial, azimuthal, and vertical velocity components, respectively. On local scales, theses cylindrical velocities have a nearly linear relation to their Cartesian counterparts, but better represent structures on scales where galactic dynamics become non-negligible (e.g., curvature of orbits). The details for the definitions of the reference frames and other parameters are largely adopted from \citetalias{ESSI}. The main differences are the switch to a new set of variables for the galactocentric frame as defined in the \textit{astropy} v4.0 frame defaults and the change of the sign for the galactocentric radial velocity $(v_R)$. A comprehensive list of the definitions, including references and a short description, is given in Appendix \ref{app:coordinates}.

% ============================================================================ %
% ============================================================================ %
\section{Methods}
\label{sec:methods}

% ============================================================================ %
\subsection{Cluster membership in the literature}
\label{sec:methods:literature}

The past few decades have seen several applications of various methods for identifying and characterizing stellar populations from astrometric data. Recent examples of techniques applied to Gaia astrometry are UPMASK \citep{upmask} and applications rooted in the popular clustering algorithms DBSCAN \citep{dbscan} and HDBSCAN \citep{hdbscan1, hdbscan2}. These algorithms have been used in extensive studies of stellar populations in the Galactic disk, for example, by \citet{cantat-gaudin18}, \citet{castro-ginard19}, and \citet{kounkel19}, respectively.

These methods identify overdensities in the five-dimensional parameter space spanned by on-sky source coordinates ($\alpha, \delta$), proper motions ($\mu_{\alpha*}, \mu_{\delta}$), and parallaxes ($\varpi)$. To some extent, photometric information is also used in the identification of distinct structures and membership determination. The applications of these algorithms have led to the identification of several hundred new open clusters in our local kiloparsec \citep[e.g.,][]{castro-ginard19, castro-ginard20}, and to the classification of numerous allegedly comoving string-like groups, see \citet{kounkel19}.

\begin{figure}
        \centering
        \resizebox{1.0\hsize}{!}{\includegraphics[]{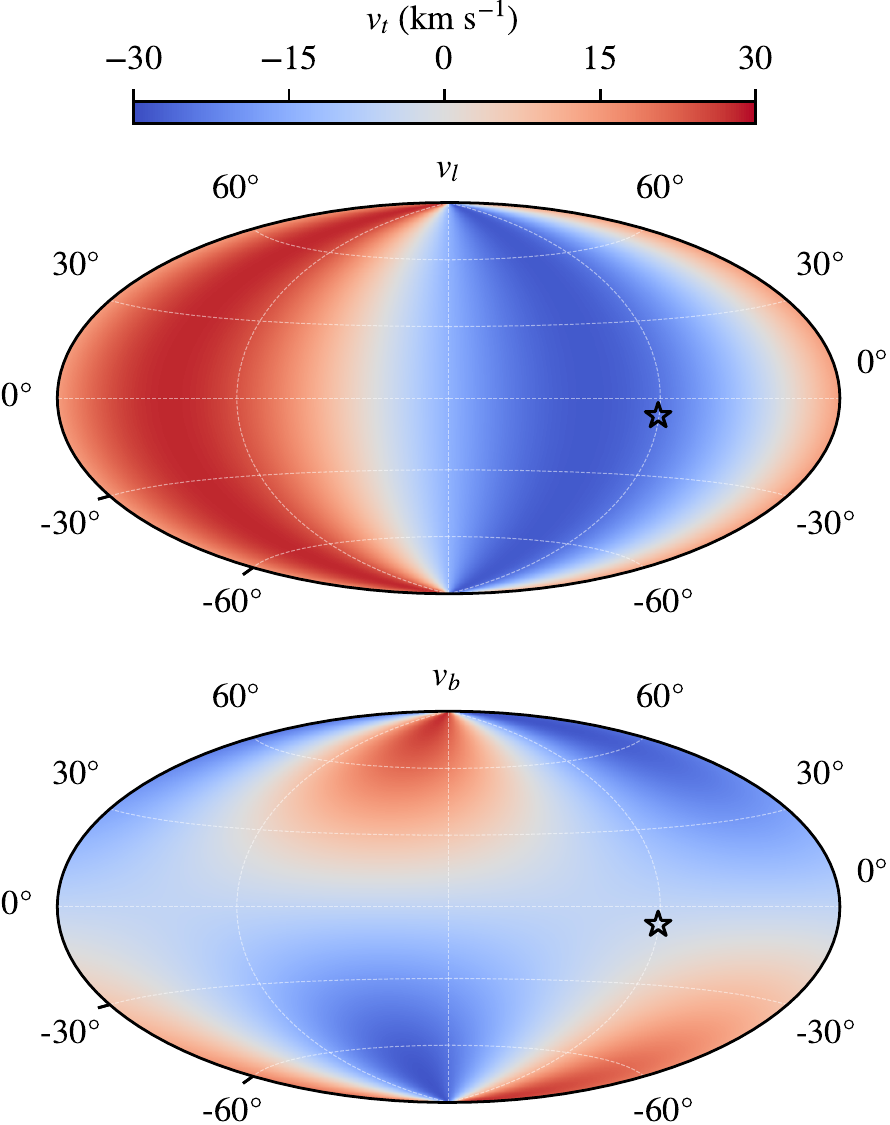}}
        \caption[]{Tangential velocity signature for sources comoving with the cluster IC~2391 in Galactic projection. The top panel displays velocities along Galactic longitude ($v_l$), and the bottom panel shows velocities along Galactic latitude ($v_b$). The position of the cluster core is marked with a black star. The tangential velocities depict a complex, highly variable pattern and illustrate the  challenge of identifying comoving populations in large regions of the sky.}
    \label{img:allsky_proper_motion}
\end{figure}

Ideally, the full six-dimensional parameters space (3D positions and 3D velocities) should be used to identify comoving stellar populations. However, this information can only be computed for the relatively small subset of sources for which radial velocity measurements are available in addition to positional and proper motion data. Conversely, identifying physically relevant structure in the five-dimensional parameter space, given by the 3D position and two tangential velocities is significantly more challenging because of projection effects across large regions ($\gtrsim$\SI{10}{\degree}) of the sky. In particular, connecting sources in proper motions (or tangential velocities) poses a delicate challenge because the proper motion signature of a comoving population (i.e., a group of stars sharing the same space motion) largely depends on our perspective. This problem is illustrated in Fig.~\ref{img:allsky_proper_motion}, which shows the expected tangential velocities of stars that are comoving with the cluster IC~2391 across the entire sky in a Galactic coordinate frame. The top and bottom panels in the figure show tangential velocities in the direction of Galactic longitude ($v_{t,l}$) and along Galactic latitude ($v_{t,b}$), respectively. Both panels show a complex pattern of tangential motions with fluctuating amplitudes and spatially inhomogeneous variations (i.e., the spatial derivatives of these patterns are also variable). Even at distances of only a few degrees from the cluster center, the measured tangential velocities can be highly variable, making comoving populations exceptionally hard to trace in tangential velocities with unsupervised clustering algorithms.

% Identifying physically relevant structures in this five-dimensional parameter space, however, is challenging due to significant projection effects across large regions ($\gtrsim$\SI{10}{\degree}) of the sky.

Because of the challenges outlined above, it is important to highlight the value of confirming and interpreting the physical significance of identified structures. Here, \gaia data directly facilitate two important measures: (a) photometry that is consistent with a single-age population in the form of a well-defined main and possibly giant sequence in observational Hertzsprung-Russel diagrams (HRD); (b) coherent 3D space motions verified by a velocity dispersion that is significantly smaller than measured for the Galactic field. In reference to point (b), star clusters in general, but also their associated tidal features in particular, are expected to feature velocity dispersions on the order of only a few \si{\km \per \second} \citep{Kuepper10a,Kuepper10b,Chumak10}. In addition, young stellar populations are also expected to feature very small velocity dispersions ($\sigma_{v_\mathrm{3D}} \lesssim$ \SI{5}{\km \per \s}; e.g., \citealp{Kuhn19}).

Velocity dispersions calculated directly from observational data are inflated by individual outliers and measurement errors (in particular from radial velocities). To remedy this problem, in all our calculations of velocity dispersions, including other published member catalogs, we first removed all sources outside a 3$\sigma$ range in any individual velocity component and then deconvolved the resulting distribution with a single 3D Gaussian \citep{xd_bovy}. Moreover, we calculated the velocity dispersion from galactocentric cylindrical components $\left( \sigma_{v_\mathrm{3D}}^2 = \sigma_{v_R}^2 + \sigma_{v_\phi}^2 + \sigma_{v_Z}^2 \right)$ to mitigate additional inflation effects resulting from curving orbits of large structures. The 3D velocity and $\sigma_{v_\mathrm{3D}}$ can only be computed from the subset of sources in a given sample that includes radial velocity data.

\begin{figure}
        \centering
        \resizebox{1.0\hsize}{!}{\includegraphics[]{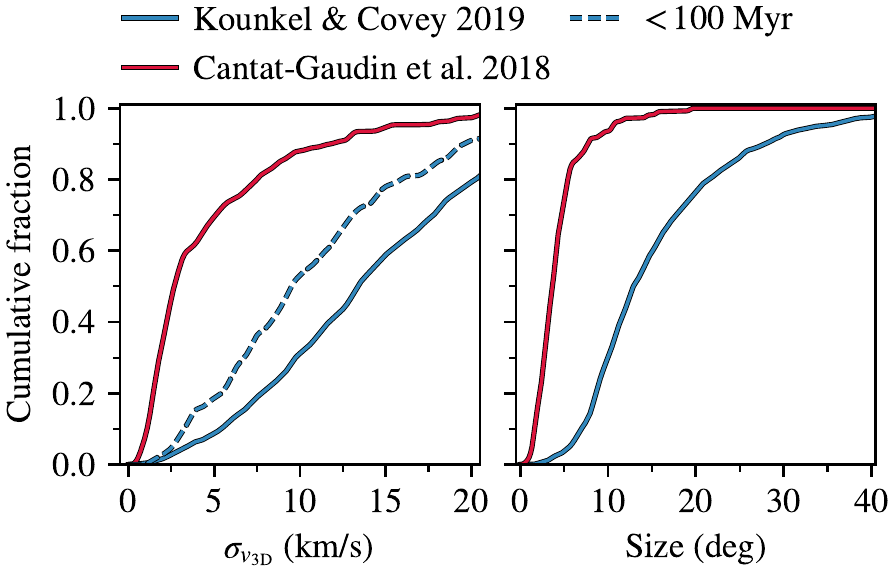}}
        \caption[]{Cumulative histograms for the deconvolved 3D velocity dispersions (left panel) and apparent sizes (right panel) of identified structures as published by \citet{cantat-gaudin18} (red) and \citet{kounkel19} (blue). Cantat-Gaudin et al. characterized members close to known clusters with relatively small velocity dispersions. Kounkel \& Covey\ identified much larger structures, but more than \SI{90}{\percent} of the populations feature velocity dispersions greater than \SI{5}{\km \per \s}. Young structures ($<$\SI{100}{Myr}) in the sample from \citet{kounkel19} (dashed blue line) also feature mostly relatively large velocity dispersions.}
    \label{img:vd_comparison}
\end{figure}

Figure~\ref{img:vd_comparison} displays the velocity dispersions (left panel) and apparent sizes (right panel) of all structures with a minimum of ten members identified in \citet{cantat-gaudin18} and \citet{kounkel19}, respectively. While \citet{cantat-gaudin18} specifically targeted known clusters for member identification, \citet{kounkel19} deployed HDBSCAN in a largely unsupervised approach. Fig.~\ref{img:vd_comparison} shows that Cantat-Gaudin et al. determined groups with overall very small velocity dispersions, where about two-thirds of the structures feature values below \SI{5}{\km \per \s}, but with relatively limited sizes. Conversely, \citet{kounkel19} established physically larger structures, but also with significantly larger (deconvolved) velocity dispersions. About \SI{90}{\percent} of the identified structures feature kinematically hot values above \SI{5}{\km \per \s}, which raises questions about the effectiveness of the authors' unsupervised approach and challenges the physical relevance of many identified structures. Furthermore, this situation changes only marginally when young groups ($<$\SI{100}{Myr}) are considered alone. The cumulative deconvolved velocity dispersion for these young groups is shown in Fig.~\ref{img:vd_comparison}. For this sample, the vast majority of structures also shows values that far exceed the expected velocity dispersions for young populations.

The particular open star clusters in our sample have been studied frequently. Nevertheless, the above-outlined results and their inherent limitations call for a more thorough treatment of the member-selection procedure. In light of the complex task of identifying connected structures in five dimensions, we introduce a new procedure in the next section that mitigates several biases in a physically motivated approach.

% ============================================================================ %
\subsection{Selecting comoving populations with tangential velocities}

\begin{figure*}
        \centering
        \resizebox{1.0\hsize}{!}{\includegraphics[]{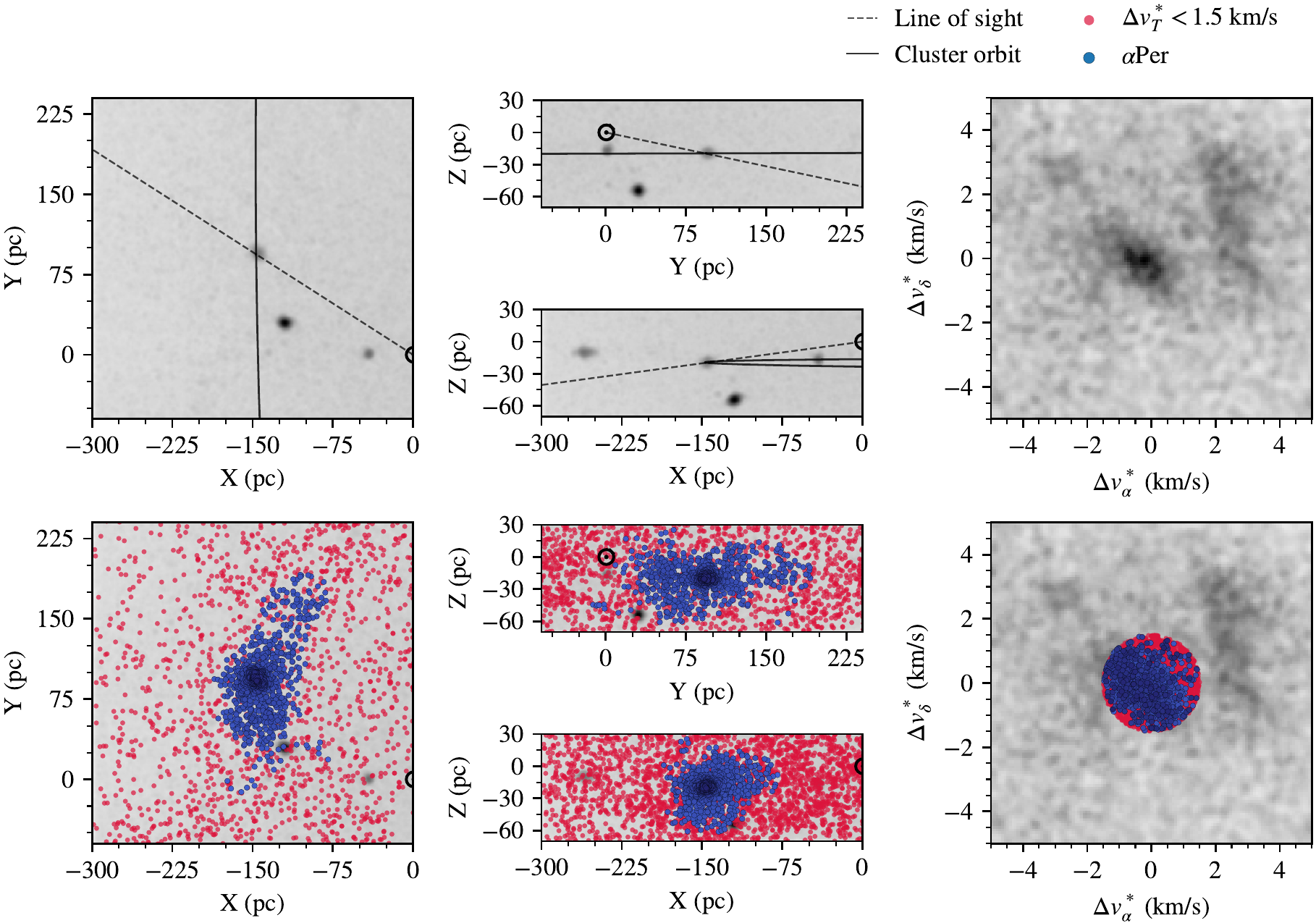}}
        \caption[]{Member selection process for the \aper cluster. The left and center columns display Galactic $XYZ$ coordinates centered on the Sun, and the right-hand side column shows deprojected tangential velocities. The top row shows kernel density maps centered on \aper that distinctly highlight the cluster populations. The line of sight (dashed line) and the Galactic orbit of \aper (solid line) are also shown. The bottom panels present the member-selection procedure, in which first a radial filter is applied in deprojected tangential velocities (red sources), followed by DBSCAN clustering in the $XYZ$ parameter space to produce the final selection (blue dots).}
    \label{img:method}
\end{figure*}

Perspective effects on tangential velocities (see Fig.~\ref{img:allsky_proper_motion}) need to be taken into account when comoving structures are identified in large regions of the sky. For this reason, we implemented a supervised method that separates spatial from kinematic dimensions and also corrects for highly variable tangential velocity signatures. The method is inspired by the popular convergent point technique \citep[e.g.,][]{convergent_point} and requires prior knowledge about the bulk 3D space motion of the target population. Our method computes for each source the difference between the actual measured tangential velocity from \gaia and the expected tangential velocity, assuming the bulk 3D velocity of a particular cluster population. In the following detailed description of the method, this difference is referred to as $\Delta v_t^* = (\Delta v_{\alpha}^{*}, \Delta v_{\delta}^{*}$) for the right ascension and declination components, respectively, and is expressed in \si{\km \per \second}. Any source that shares the exact same space velocity with the input vector (the cluster bulk velocity) therefore fulfills $\Delta v_{\alpha}^{*} = \Delta v_{\delta}^{*} = 0$ \SI{}{\km \per \s}. Sources with marginally different 3D velocities feature slightly larger offsets to the origin in this transformed parameter space. As a result, our method facilitates efficient filtering of kinematically incompatible sources and can be easily followed-up with well-established clustering methods in the $XYZ$ space.

As input, our supervised method requires establishing the bulk cluster velocity at a given 3D space coordinate. We used the cluster membership tables published by \citet{cantat-gaudin18}, where we retained only sources with membership probabilities greater than 0.8. We then computed mean positions and velocities after applying a 3$\sigma$ clipping iteratively for five times. The mean velocities in galactocentric cylindrical coordinates for each cluster are listed in Table~\ref{tab:clusters}. Based on the calculated mean $XYZ$ position, we sliced the input catalog (see Sect.~\ref{sec:data}) to retain only sources in a generously large box with edge lengths of \SI{300}{pc} in $X$, \SI{500}{pc} in $Y$, and \SI{100}{pc} in $Z$, centered on the 3D cluster positions determined before. These limits were determined iteratively and were chosen as a compromise between not confining our selection by these limits and reducing contamination in the transformed tangential velocity space. Within our already stringent constraints on data quality, this procedure typically leaves several hundred thousand sources in the sliced box of each cluster.

The next paragraphs outline the method behind our membership determinations in detail. Figure~\ref{img:method} illustrates all steps in the selection procedure in reference to the \aper group and serves mainly as a guidance for our explanations. The left and central columns show the spatial arrangement of sources in Galactic Cartesian coordinates, and the column on the right displays kinematic information. The panels in the top row feature kernel density maps of the sliced box around the cluster and furthermore indicate the line of sight and the cluster Galactic orbit. The $XYZ$ and $\Delta v_{\alpha}^{*}$-$\Delta v_{\delta}^{*}$ maps were constructed with Epanechnikov kernels with sizes of \SI{5}{pc} and \SI{0.25}{\km \per \s}, respectively. These particular kernel sizes are not physically motivated, but instead were chosen for visualization purposes. Already in the 3D position data, \aper appears as an easily discernible overdensity at the center of the panels, together with other prominent clusters (in this case, the Pleiades and Hyades). Similarly, the transformed velocity space in the right-hand side panel also features a well-defined overdensity near the origin, depicting sources with a kinematic profile similar to that of \aper.

In the second step, our procedure filters sources that are kinematically incompatible with the cluster bulk motion. To this end, we experimented with several methods, ranging from simple thresholding to other popular clustering methods such as DBSCAN and OPTICS \citep{optics}. However, for several of our target clusters (and also associations and clusters outside of our sample), the distribution in the $\Delta v_{t}^{*}$ plane did not resemble such clearly separable overdensities as in the case of \aper in the upper right panel of Fig.~\ref{img:method}. As a consequence, the algorithms performed poorly in many cases, and we instead chose to apply a hard limit of \SI{1.5}{\km \per \s} about the cluster bulk motion. This step drastically reduced the number of remaining sources by more than two orders of magnitude to a few thousand kinematically compatible sources. This reduced selection is displayed in the bottom panels of Fig.~\ref{img:method}.

In a third and last step we extracted overdensities in the $XYZ$ space from the kinematically filtered sample. To this end, we used DBSCAN with $\epsilon =$\SI{15}{pc} and \textit{minPts} = 5. Furthermore, we only retained core points from the DBSCAN algorithm in order to keep the contamination at the population perimeters as low as possible. We experimented with several parameter combinations, and with the aid of the resulting observational HRDs, again chose a compromise between a too confined selection (only cluster core) and too many contaminating field stars. The resulting final selection for \aper is displayed in the bottom panels of Fig.~\ref{img:method}. For most of our target clusters, this procedure resulted in a single, clearly separable population. Only in very few cases did our DBSCAN setup also label other (assumed to be unrelated) populations, which we chose to not investigate further here.

The choice of the free parameters for our method (filter radius in $\Delta v_{t}^{*}$, DBSCAN $\epsilon,$ and \textit{minPts}) is accompanied by a set of caveats. For instance, the hard clipping in the transformed tangential velocity space only represents a subset of the kinematically compatible population because members may very well lie beyond this radius. Moreover, in reference to the top right panel in Fig.~\ref{img:method}, the overdensity in velocities is clearly not circular and seems to extend beyond the \SI{1.5}{\km \per \s} limit. The overdensity instead appears to take an elliptical shape, which is also confirmed in the final selection in the bottom right panel. Other disadvantages relate to the fixed parameter setup for DBSCAN when searching for overdensities in XYZ. For instance, all field stars that by chance share a similar velocity and are located in the same volume are selected as well. Moreover, measurement errors in geometric distances along the line of sight result in elongated structures, thereby distorting the $XYZ$ space. This effect depends on the measurement errors in geometric distance and is more pronounced for more remote populations. Taken together, our parameter choice was optimized for the  target clusters and should be carefully evaluated when this method is extended to other populations.

% ============================================================================ %
\subsection{Deconvolution of the spatial distribution}

\begin{figure}
        \centering
        \resizebox{1.0\hsize}{!}{\includegraphics[]{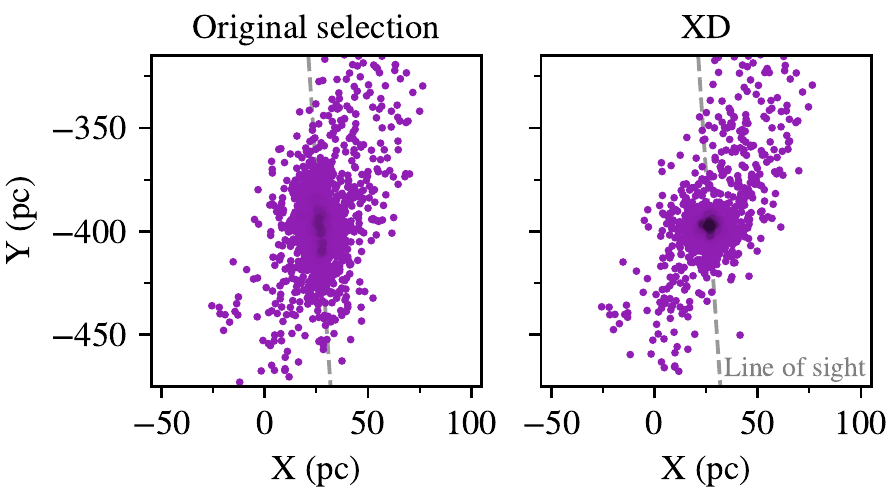}}
        \caption[]{Spatial configuration of the central regions of NGC~2516 before (left) and after (right) deconvolving source distances. The original distribution clearly shows a substantial elongation effect along the line of sight (dotted gray line), while the extreme deconvolution (XD) infers the expected spherical arrangement of the cluster core.}
    \label{img:method_xd}
\end{figure}

The spatial arrangement of our selections for the ten target clusters appears distorted through significant measurement errors in distances along the line of sight. For example, for the most remote cluster population in our sample, NGC~2516, the cluster core is at a distance of about \SI{410}{pc}. Even with our rigorous quality criteria, we expect standard errors in distance up to \SI{40}{pc}. With typical cluster core sizes of only a few parsec, we consequently observe significant elongation effects in the cluster appearance. However, for reliable determinations of physical cluster sizes and cluster kinematics, for instance, this effect needs to be accounted for.

To infer the underlying true spatial arrangement of the populations, we employed a novel algorithm that is described in detail in \textcolor{blue}{Meingast 2020 (in prep.)}. Here, we provide a brief summary of its functionality. In a first step, the algorithm constructs covariance matrices for each source with respect to the measurement errors in $XYZ$ positions. This is done by randomly sampling the observed parallax from the (symmetric) error distribution. Errors in on-sky position ($\alpha, \delta$) are ignored because they are negligibly small. The $XYZ$ distribution, informed by the covariance matrices of the given spatial arrangement, is then deconvolved with a mixture of Gaussians, a process referred to as \textup{\textup{\textit{\textup{extreme deconvolution}}}} in the literature \citep[XD;][]{xd_bovy}. The dimensionality of the mixture model (i.e., the number of Gaussian components) is determined with the Bayesian information criterion \citep{bic}. The number of components for our target clusters is typically between 3 and 5.

This step results in the description of the spatial configuration of a population in the form of a probability distribution composed of a mixture of 3D Gaussians. In a second step, the method infers the geometric distance of each source along the line of sight by using the mixture model as distance prior together with a Gaussian likelihood from the parallax $\varpi$ measurement by \gaia. To obtain the deconvolved distances, we sampled the posterior with the Markov chain Monte Carlo ensemble sampler (MCMC) \textit{emcee} \citep{emcee1, emcee2}.

The results of this procedure are visualized in Fig.~\ref{img:method_xd}. The figure shows the $XY$ distribution of the central parts of NGC~2516, where the luminance value of the points is scaled proportionally to volume density. The left panel displays the positions when the \gaia parallaxes are directly inverted. In this view, the cluster appears highly elongated along the line of sight. The panel on the right-hand side displays the median posterior positions in the MCMC sampling. Our method clearly infers a good approximation of the true shape of the cluster core.

% ============================================================================ %
\subsection{Contamination}
\label{sec:methods:contamination}

\begin{figure}
        \centering
        \resizebox{1.0\hsize}{!}{\includegraphics[]{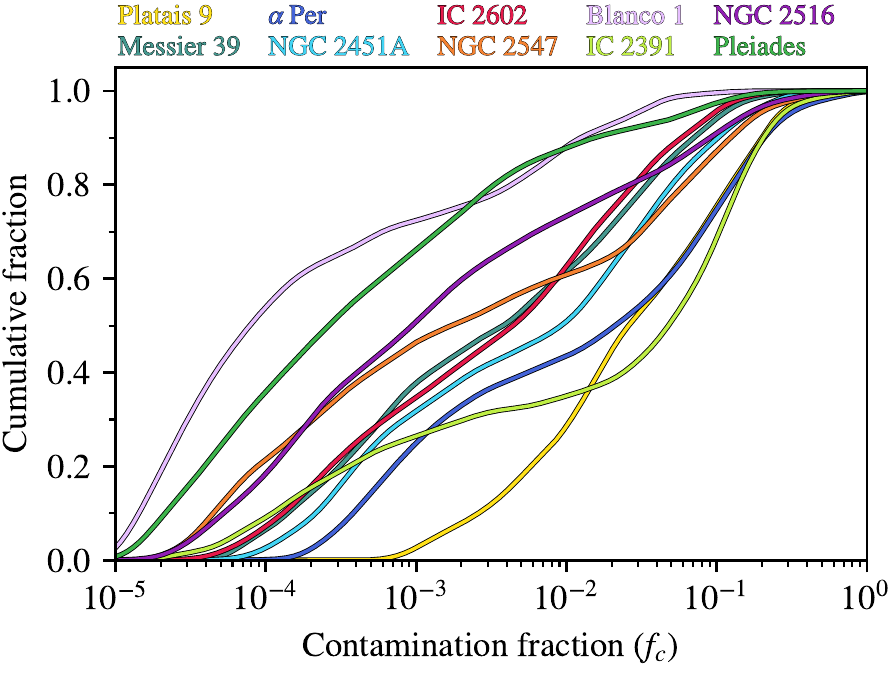}}
        \caption[]{Cumulative distributions of the source contamination fractions $f_c$ for each cluster. The contamination fraction $f_c$ denotes the volume density contrast for each source with respect to the surrounding background population.}
    \label{img:contamination}
\end{figure}

In this method, the tangential velocity filter effectively removes most field star contaminants. However, all sources that pass this filter and are clustered with the subsequent application of DBSCAN will become part of the final member selection. In this way, some stars that are part of the Galactic field or other populations which, by chance, share similar tangential velocities, will inevitably also contaminate our population selection. In the following description, we refer to these contaminating sources as background population.

To estimate the number of contaminating background sources, we exploited the specific workflow of our method. To this end, we compared the finally selected cluster population with all remaining sources after the initial constraint in tangential velocities  (i.e., we compared the red and blue sources in Fig.~\ref{img:method}). We then obtained a contamination estimate by comparing stellar volume densities of our extracted cluster populations $\rho_{\mathrm{cl}}$ to the median volume density of all remaining background sources $\langle \rho_{\mathrm{bg}} \rangle$ in the corresponding \gaia subset. Specifically, we parameterized the volume density at the location of each source via the distance to its seventh nearest neighbor and chose the median background density as comparison to avoid edge effects. With these two parameters at hand, we obtained fractional contamination estimates for each source ($f_{c,i}$) via

\begin{equation}
    f_{c,i} = \frac{\langle \rho_{\mathrm{bg}} \rangle}{\rho_{\mathrm{cl,i}} + \langle \rho_{\mathrm{bg}} \rangle}.
\end{equation}

Following this definition, the parameter $f_c$ can be viewed as a measure of the contrast between cluster and background stars. For example, $f_c$ takes a value of 0.5 for a cluster member located at a position with equal cluster and background volume densities. On the other hand, $f_c$ is close to 0 for sources that are surrounded mostly by their population siblings, while rogue cluster members will have values close to 1.

Figure~\ref{img:contamination} displays the cumulative distribution of $f_c$ for each population. The figure reveals that the large majority of the determined cluster members show exceptionally low values of fractional contamination. Typically, the majority of sources is located at positions where members outnumber background stars by a factor of 100 or more ($f_c < 10^{-2}$). The fractional contamination increases toward the outskirts of a population, whereas the lowest values for $f_c$ are found at the center of the most compact clusters (Blanco~1, Pleiades, and NGC~2516). Blanco~1 is characterized overall by very low values of $f_c$, which can be attributed to its location about \SI{200}{pc} below the Galactic plane, where the field star population is already thinned out. This is also illustrated in Fig.~\ref{img:contamination_xy}, which shows the source distribution for each cluster in the $XY$ plane, color-coded with individual contamination estimates.

\begin{table*}[!t]
\caption{Computed cluster parameters and statistics. The columns list a variety of calculated parameters, where $d_c$ is the redetermined distance to the population density maxima, N is the number of selected sources, N$_c$ is the estimated total number of contaminants, M$_{\mathrm{tot}}$ is the cumulative total mass of the observed sources, M$_{\mathrm{tot}}^*$ is the total system mass based on an extrapolation with mass functions, $f_\mathrm{M,t}$ is the mass fraction inside the tidal radius of a system, $r_t$ and $r_h$ are the tidal and half-mass radii, respectively, $\sigma_{v_\mathrm{3D}}$ is the deconvolved 3D velocity dispersion in galactocentric cylindrical coordinates, and $k_\mathrm{xu}$, $k_\mathrm{yv}$, and $k_\mathrm{zw}$ are expansion rates and their statistical errors in the corresponding position-velocity planes.}
\begin{tabular*}{\linewidth}{@{\extracolsep{\fill}} l c c c c c c c c c c c c}
\hline\hline
Cluster & $d_c$ & N (N$_c$) & M$_{\mathrm{tot}}$ & M$_{\mathrm{tot}}^*$ & $f_\mathrm{M,t}$ & $r_t$ & $r_h$ & $\sigma_{v_\mathrm{3D}}$ & $k_\mathrm{xu}$ & $k_\mathrm{yv}$ & $k_\mathrm{zw}$ \\
& (pc) &  & (M$_\sun$) & (M$_\sun$) &  & (pc) & (pc) & (\si{\km\per\second}) & (\si{\metre\per\second\per pc}) & (\si{\metre\per\second\per pc}) & (\si{\metre\per\second\per pc}) \\
\hline
Platais 9 & 185.4 & 328 (22) & 187.2 & 285.0 & 0.07 & 3.6 & 29.5 & 3.6 & 31.7 $\pm$ 8.3 & 13.8 $\pm$ 21.1 & 25.3 $\pm$ 21.0 \\
Messier 39 & 297.6 & 351 (7) & 235.2 & 325.0 & 0.40 & 6.7 & 9.9 & 0.5 & 32.5 $\pm$ 9.8 & 6.0 $\pm$ 18.6 & -61.5 $\pm$ 33.5 \\
\aper & 175.0 & 1223 (91) & 735.0 & 1030.0 & 0.41 & 10.0 & 16.4 & 3.8 & 39.1 $\pm$ 18.4 & 14.5 $\pm$ 11.5 & 19.3 $\pm$ 5.9 \\
NGC 2451A & 193.8 & 648 (21) & 327.1 & 425.0 & 0.39 & 7.2 & 11.3 & 3.3 & 37.6 $\pm$ 3.9 & 44.3 $\pm$ 14.4 & 23.3 $\pm$ 10.9 \\
IC 2602 & 152.2 & 648 (12) & 339.7 & 400.0 & 0.47 & 7.5 & 8.3 & 1.1 & 42.1 $\pm$ 5.8 & -26.3 $\pm$ 22.9 & -44.3 $\pm$ 14.3 \\
NGC 2547 & 392.0 & 514 (20) & 382.8 & 590.0 & 0.52 & 8.9 & 8.1 & 1.3 & 11.6 $\pm$ 9.7 & 6.7 $\pm$ 35.3 & -12.8 $\pm$ 24.1 \\
Blanco 1 & 237.5 & 494 (2) & 259.9 & 365.0 & 0.76 & 8.5 & 3.9 & 0.7 & 47.7 $\pm$ 22.2 & 4.7 $\pm$ 3.1 & 103.5 $\pm$ 72.8 \\
IC 2391 & 152.1 & 682 (55) & 338.4 & 445.0 & 0.37 & 7.2 & 23.8 & 2.9 & 17.1 $\pm$ 8.0 & -3.8 $\pm$ 12.3 & 17.6 $\pm$ 17.3 \\
NGC 2516 & 413.8 & 1860 (51) & 1436.7 & 2550.0 & 0.71 & 16.0 & 8.9 & 1.4 & 44.5 $\pm$ 5.1 & 22.8 $\pm$ 6.3 & 27.2 $\pm$ 13.9 \\
Pleiades & 136.4 & 1177 (9) & 617.8 & 850.0 & 0.82 & 11.8 & 4.4 & 1.4 & 0.9 $\pm$ 21.5 & 7.9 $\pm$ 7.4 & -9.2 $\pm$ 19.6 \\
\hline
\end{tabular*}
\label{tab:clusters_observables}
\end{table*}

In addition, the parameter $f_c$ also enables an estimate of the total contamination by summing over all fractional contamination measures for each source. The measured total contamination by background stars amounts to well below \SI{10}{\percent} for all populations, with a maximum of \SI{8}{\percent} for IC~2391, values as low as \SI{1}{\percent} for the Pleiades and Blanco~1, and a median of \SI{3}{\percent} across all ten target clusters. The estimated number of total contaminating sources is listed in Table~\ref{tab:clusters_observables} for each population, and $f_c$ values for each source are available online (Table~\ref{tab:sources}).

% ============================================================================ %
% ============================================================================ %
\section{Results}
\label{sec:results}

This section contains a presentation of our results from an observational point of view. All results presented in this section are based on our application of the methods outlined in Sect.~\ref{sec:methods} to the ten target clusters listed in Table~\ref{tab:clusters}. This procedure includes rigorous filters that are used to retain only sources that are both kinematically compatible with the cluster bulk motion and are spatially connected to the cluster cores. Moreover, we also minimized the effect of structure elongation due to large measurement errors in the geometric distances along the line of sight. We describe subjects including the observational HRDs of the clusters (Sect.~\ref{sec:results:hrd}), their spatial distribution (Sect.~\ref{sec:results:spatial}), present-day mass functions (Sect.~\ref{sec:results:massfunc}), radial profiles (Sect.~\ref{sec:results:profiles}), and cluster kinematics (Sect.~\ref{sec:results:kinematics}). Each individual topic is discussed only broadly, and our results should serve as a presentation of the observables. 

% Removed as per referee request
% As such we are not aiming for completeness in our analysis as our findings would certainly warrant a much more detailed investigation for each individual population. We also refrain from presenting the results on each group individually to keep this manuscript at a reasonable volume and rather discuss their properties in a global context. Table~\ref{tab:clusters_observables} lists all discussed values and is referred to multiple times throughout this section.

% ============================================================================ %
\subsection{Observational Hertzsprung-Russel diagrams}
\label{sec:results:hrd}

\begin{figure*}[!ht]
        \centering
        \resizebox{1.0\hsize}{!}{\includegraphics[]{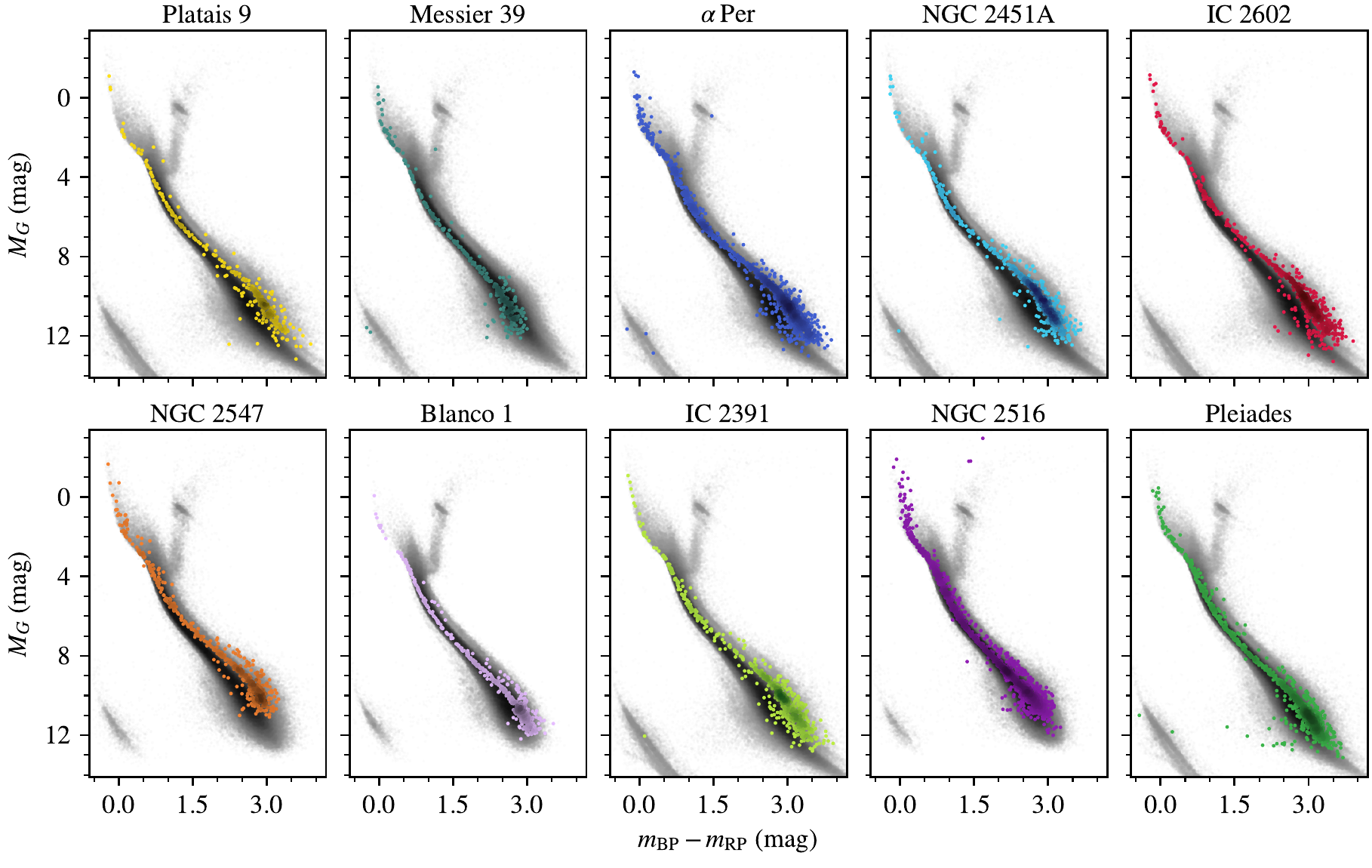}}
        \caption[]{Observational HRDs for the ten target clusters. Members of each cluster are displayed as colored points. The gray distribution in the background comprises all \gaia sources in the initial search box of each population. Our selection results in remarkably clean and narrow main sequences that frequently also show well-visible binary sequences. This clear separation from the broad background distribution acts as validation of the successful application of our member-selection procedure.}
    \label{img:cmds}
\end{figure*}

At the start of our investigation, we created observational HRDs for the member selection of each target cluster. In these diagrams, the main sequence of a single-age population appears as a narrow distribution, while a random selection would result in more broadly scattered absolute magnitudes and colors. We therefore used the HRDs mainly as a diagnostic tool to validate our selection and to show that the extracted source populations are likely coeval.

Figure~\ref{img:cmds} displays the HRDs of the ten target clusters, where each cluster is displayed in its respective color in addition to all sources in the initial \SI{300 x 500 x 100}{pc} search box of each group. The diagrams show that the populations appear as exceptionally narrow main sequences. Likewise, the sequences appear in stark contrast to their respective background population. This observation is further confirmed by performing a two-dimensional version of the Kolmogorov-Smirnov test \citep{Peacock83} between the determined cluster members and all sources in their respective search boxes in the color-absolute magnitude space. These tests deliver p-values that are equivalent to 0 with respect to the floating-point precision. The same test on an equal-sized random draw of the background sample typically results in $p>0.5$. Another indicator for a clean selection procedure are the frequently well-visible stellar multiple sequences, even down to the lowest stellar masses. Without further consideration, we take note that especially NGC~2516 shows a remarkably distinct stellar multiple sequence. Following these findings and also including the low contamination rate as detailed in Sect.~\ref{sec:methods:contamination}, we conclude that our procedure indeed extracts coeval populations.

For Messier~39, \aper, and NGC~2516, we also find giant stars in their respective HRDs. The location of the sources on the giant branch for Messier~39 and \aper do not match an expected evolutionary track for these clusters, which likely makes them contaminants. The three sources on the giant branch in the NGC~2516 selection are \mbox{\object{HD 64320}}, \mbox{\object{HD 65662}}, and \mbox{\object{CD-60 1965}}. All of them are classified as giants \citep{giants1, giants2, giants3} and match an isochrone for the cluster age as listed in Table~\ref{tab:clusters}. Moreover, some selections also contain white dwarfs, with particular objects of interest in the very young clusters NGC~2451A and IC~2391. If these sources are part of the population, their progenitors must have had a mass similar to early B-type stars to already have produced white dwarfs.

\begin{figure*}
        \centering
        \resizebox{1.0\hsize}{!}{\includegraphics[]{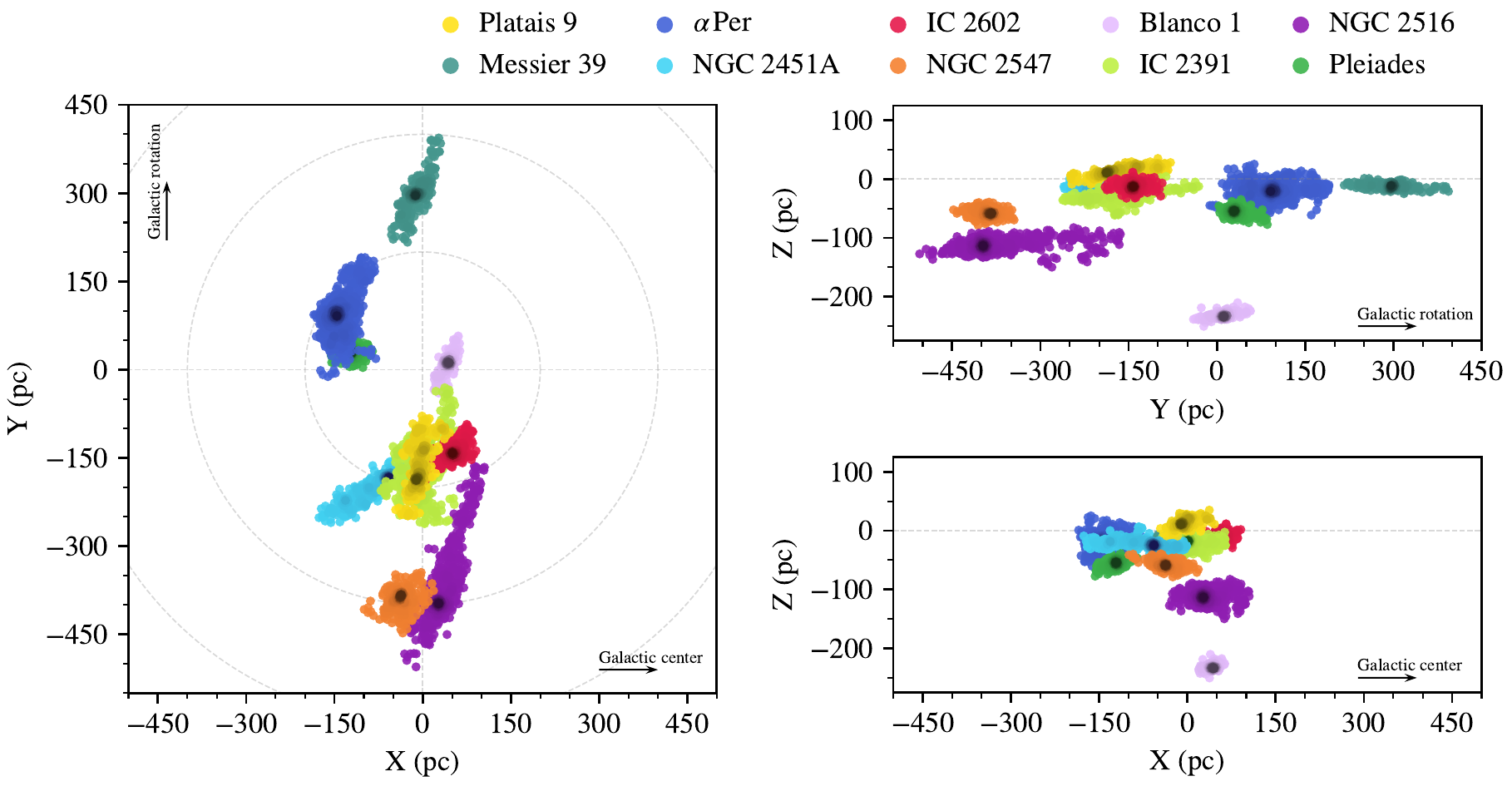}}
        \caption[]{Spatial distribution of our selection for the ten clusters in heliocentric Galactic coordinates. The color scale is proportional to volume density, accentuating the dense central regions of the populations. In this view, the Sun is located at the origin, and the directions toward the Galactic center (positive $X$) and of the generic Galactic rotation (positive $Y$) are indicated with arrows. Cluster cores are surrounded by enormous stellar coronae with sizes up to several hundred parsec. In addition, we observe a characteristic pattern where extended cluster populations form elongated shapes whose leading arm is tilted toward the inner Galaxy. An interactive 3D version of this figure is also available online at \href{https://homepage.univie.ac.at/stefan.meingast/coronae.html}{https://homepage.univie.ac.at/stefan.meingast/coronae.html}.}
    \label{img:xyz}
\end{figure*}

We furthermore find several sources located between the main sequence and the white dwarf locus. These are particularly well visible at the faint end of the selection for the Pleiades. A closer examination reveals that these sources are almost exclusively characterized by an above-average BP/RP flux excess, indicating issues with the photometry. When the blue band is replaced with the $G$ band in the HRDs (i.e., $m_G - m_\mathrm{RP}$ are plotted on the abscissa), this problem is mostly solved, which further indicates problems with the BP photometry for the few affected sources. We note here that our selection method does not include any photometric criteria, which means that these few scattered sources are likely not contaminants.

% ============================================================================ %
\subsection{Spatial distribution}
\label{sec:results:spatial}

Figure~\ref{img:xyz} shows the spatial arrangement of the extracted populations in a Galactic Cartesian coordinate frame centered on the Sun\footnote{An interactive 3D visualization of the spatial arrangement as visible in Fig.~\ref{img:xyz} of the populations is available online at \href{https://homepage.univie.ac.at/stefan.meingast/coronae.html}{https://homepage.univie.ac.at/stefan.meingast/coronae.html}, also including a comparison to the catalogs from \citet{cantat-gaudin18} and \citet{kounkel19}. In addition, Fig.~\ref{img:wcs_matrix} in the appendix shows the on-sky distribution of the populations.}. For the purpose of highlighting the cluster cores, the lightness values of the colors represent volume density, with higher densities corresponding to lower lightness values. Evidently, the cluster populations extend far beyond their long-known central regions and are surrounded by vast stellar halos we term coronae, reaching sizes up to several hundred parsec in diameter. As a result, the cluster cores, together with their massive coronae, form large, coeval, and comoving extended cluster populations, each encompassing up to hundreds of thousands of cubic parsec of space. Moreover, the groups appear to be spatially intertwined, where NGC~2451A, IC~2391, and IC~2602, and \aper and the Pleiades share a considerable volume in which the populations mix.

A closer inspection of the XY plane in Fig.~\ref{img:xyz} shows that the cluster populations altogether form highly elongated shapes, where the parts closer to the Galactic center form leading arms in the direction of Galactic rotation. In contrast, the sections located farther away from the Galactic center appear to lag behind. We also observe a variety of position angles of the coronae with respect to their orientation in the Galactic plane. While NGC~2451A, for example, is visibly more tilted toward the Galactic center, most of the other groups appear to have a similar (but not exactly the same) orientation, being more elongated in the direction of Galactic rotation. We note that the alignment of the coronae is clearly not correlated with the line of sight from our vantage point, an artificial effect that often originates in problems with respect to significantly enhanced errors in parallax or poor method deployment.

For most populations, we observe a correlation between vertical tilt (best visible in the top right panel in Fig.~\ref{img:xyz}) and vertical velocity (listed in Table~\ref{tab:clusters}). In the groups with positive vertical velocity, the leading parts of Platais~9, \aper, IC~2391, and NGC~2516 are also oriented toward the Galactic north pole. Messier~39, NGC2451~A, NGC~2547, and the Pleiades have negative vertical velocity, and their leading parts are tilted toward the Galactic south pole. IC~2602 has the highest vertical velocity, but does not show a similar trend, and interestingly, the situation is reversed for Blanco~1 (negative vertical velocity, leading part tilted toward the Galactic north pole). It therefore appears that the groups are preferentially oriented toward their direction of motion, but this does not seem to be a universal occurrence.

\begin{figure*}
        \centering
        \resizebox{1.0\hsize}{!}{\includegraphics[]{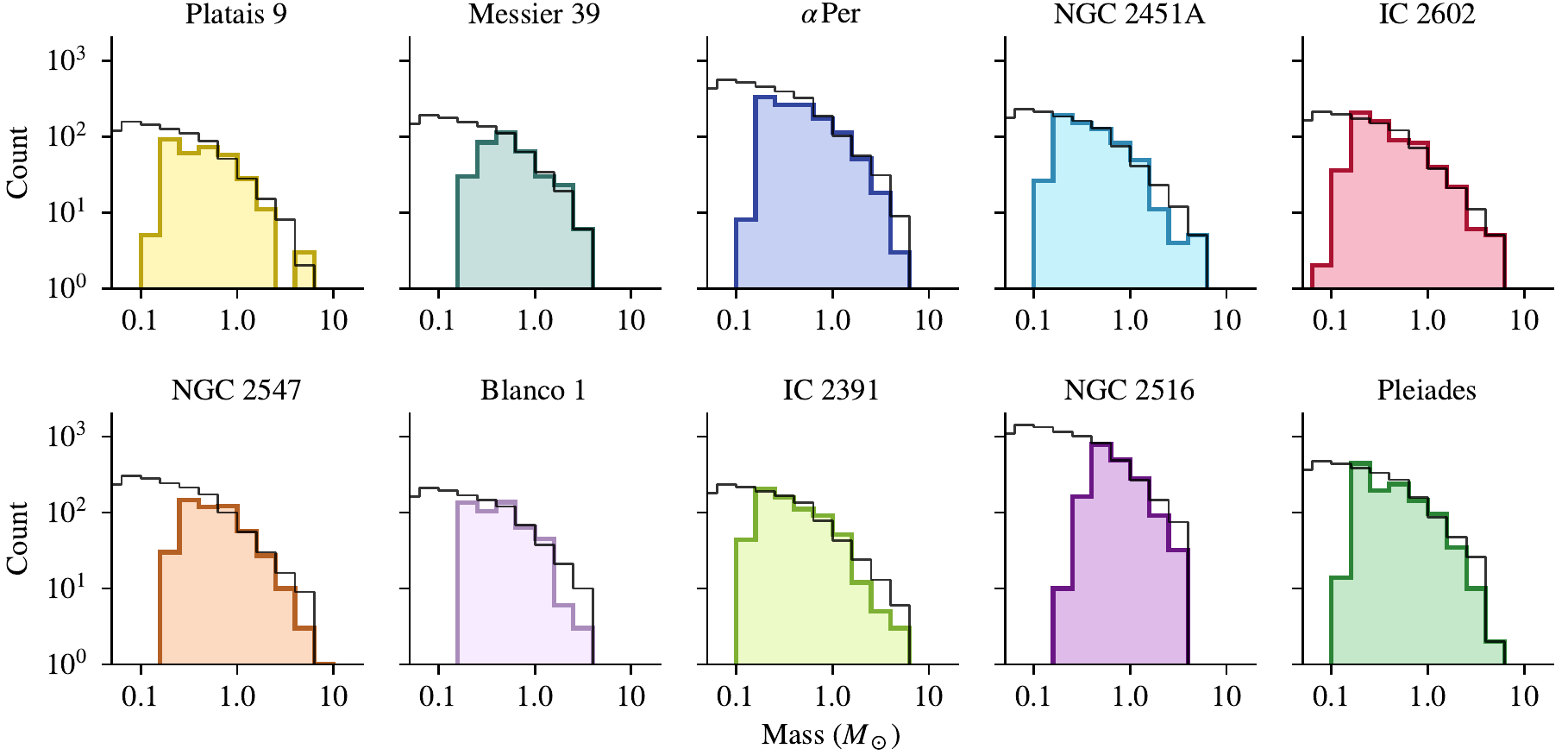}}
        \caption[]{Present-day mass functions (colored histograms) and the best-fitting truncated IMFs (solid black histogram) for each population. The completeness limits on the low-mass end appear different for each cluster because of the distance-dependent sensitivity limits. The completeness at the high-mass end is more difficult to evaluate because individual stellar evolutionary histories are governed by the independent ages of the groups. If the mass distribution follows the displayed truncated standard IMF, we estimate the missing mass fraction to be in a range from about \SI{15}{\percent} for IC~2602 to almost \SI{45}{\percent} for NGC~2516.}
    \label{img:mass_functions}
\end{figure*}

Following the deconvolution procedure, we also redetermined the 3D position of the population density maxima (i.e., the cluster cores) from the Gaussian mixture model. These new distances are listed in Table~\ref{tab:clusters_observables}. With reference to Fig.~\ref{img:xyz} we find that the cluster cores seem to be preferentially located toward the center of the populations. Only NGC~2516 seems to be an exception: the trailing arm of the corona appears less prominent than the leading part. We attribute this to observational limits and our constraints on data properties, however, that significantly thin out source densities at distances beyond \SI{{\sim}400}{pc}, which is the distance to the cluster core of NGC~2516.

% ============================================================================ %
\subsection{Mass functions}
\label{sec:results:massfunc}

In this section, we derive total mass estimates for each population in order to lay the foundation for subsequently establishing physical properties. Determining the total mass requires an extrapolation for low-mass sources because they are beyond a distance-dependent observational completeness limit.

We first obtained individual masses for each observed source by interpolating their absolute magnitudes and colors on a given isochrone. To this end, we used PARSEC isochrones \citep{Bressan12} with the \citet{Weiler18} revised Gaia DR2 passbands and adopted ages and (assumed to be constant) line-of-sight extinctions from \citet{Bossini19}. Figure~\ref{img:mass_functions} shows the resulting measured mass functions for each target population as colored histograms. As expected, sensitivity limits cause a truncation at the low-mass end that is particularly well visible for NGC~2516, the most remote population in our sample. Different observed maxima at the high-mass end are also visible, where some populations show a well-sampled distribution up to several solar masses, while others (e.g., NGC~2516) are apparently truncated. In contrast to low-mass stars that can drop out of our selection because of sensitivity limits, the high-mass range can be affected by both incomplete catalog coverage (too bright sources) and stellar evolution processes (supernovae in the past).

To facilitate an extrapolation to total present-day cluster masses, we compared the measured distributions to \citet{Kroupa01} initial mass functions (IMF). Specifically, we decided to sample IMFs truncated to the highest measured stellar mass because in our selection we can hardly distinguish between mass loss due to stellar evolution and an observational bias in the \gaia database. The limit on the low-mass end was set to \SI{0.03}{M_\sun}, thus also accounting for potentially missing mass below the hydrogen-burning limit. We then obtained the best-fit mass function by minimizing the total mass difference between a given (truncated) IMF and the measured total masses. Because of completeness limits in the Gaia data, only the mass range between 0.5 and \SI{2}{M_\sun} was used for this comparison. The lower limit of \SI{0.5}{M_\sun} corresponds to an M1 dwarf with an absolute $G$ magnitude of about \SI{8.8}{mag} \citep{Pecaut13,Evans18}. For a distance modulus of $\mu=8.5\,\si{mag}$ (i.e., a distance of \SI{500}{pc}), this corresponds to an apparent $G$ magnitude of \SI{17.3}{mag} for the lower mass limit, which is well within the Gaia completeness threshold \citep{Arenou18} and even leaves an ample margin for interstellar extinction. The resulting best-fit mass functions are displayed in Fig.~\ref{img:mass_functions}. The measured total mass M$_{\mathrm{tot}}$ and the total mass of the best-fit mass function M$_{\mathrm{tot}}^*$ are listed in Table~\ref{tab:clusters_observables}. The missing mass fraction due to incompleteness at the low-mass end is between \SI{15}{\percent} (IC~2602) and \SI{45}{\percent} (NGC~2516), with a mean of about \SI{30}{\percent} across all ten clusters.

% ============================================================================ %
\subsection{Radial profiles}
\label{sec:results:profiles}

\begin{figure*}
        \centering
        \resizebox{1.0\hsize}{!}{\includegraphics[]{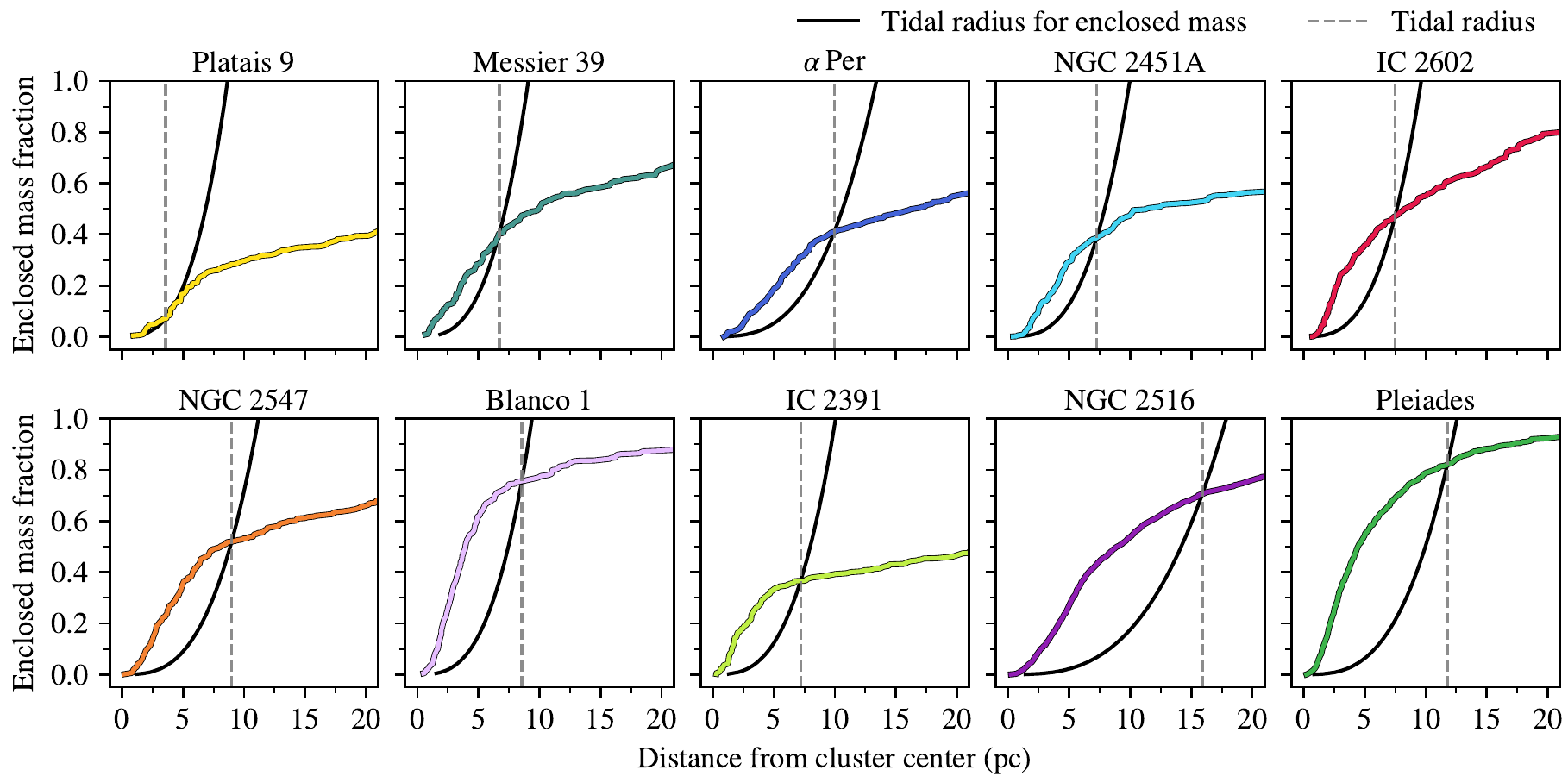}}
        \caption[]{Radial mass profiles and tidal radii for each cluster population. The colored lines depict the enclosed mass fraction as a function of distance from the density maximum of each population. The solid black lines trace the tidal radius for a given enclosed total mass content. The intersection of these profiles identifies the tidal radius. The mass profiles themselves depict a variety of shapes, with some clusters showing clear kinks near the tidal radius. For the majority of the clusters, the mass profiles also reveal that the bulk of their stellar mass content is located beyond the tidal radii in the stellar coronae.}
    \label{img:tidal_radii}
\end{figure*}

Our completeness-corrected mass estimates, together with the deconvolved spatial distribution, enabled a determination of physically relevant geometric parameters, specifically, tidal and half-mass radii ($r_t$ and $r_h$). For our purpose, we use the terms of the Jacobi and tidal radii interchangeably \citep[cf.][]{Binney08} and follow the relation outlined by \citet{King62} and \citet{Ernst10} with
\begin{equation}
    r_t = \left[ \frac{GM}{\left( 4 - \beta^2 \right) \Omega^2} \right]^{1/3},
\end{equation}
where $G$ is the gravitational constant, $M$ is the enclosed total mass, $\Omega$ is the circular frequency of a near-circular orbit at the distance of the population to the Galactic center, and $\beta = \kappa / \Omega$, with $\kappa$ depicting the epicyclic frequency. Here, we calculated the circular and epicyclic frequencies for each population with the \code{MWPotential2014} potential as defined in \textit{galpy} \citep{galpy}.

We continued to calculate cumulative enclosed total masses as a function of distance from the density maximum of each population. To account for sources beyond the sensitivity limit of \gaia, the measured enclosed total masses were multiplied by the individual incompleteness factors as determined from the mass functions in Sect.~\ref{sec:results:massfunc}. At the same time and using the above equation, we also calculated tidal radii for the enclosed mass at a given radial distance to the cluster center. This procedure provided tidal radii as a function of distance from the population centers (or conversely, as a function of enclosed mass). The intersection of the (completeness-corrected) radial mass profile and the tidal radius profile then yielded the actual tidal radius for each system. The values for the determined tidal radii, along with the system half-mass radii, are reported in Table~\ref{tab:clusters_observables}. Figure~\ref{img:tidal_radii} illustrates this procedure. The measured tidal radius (i.e., the intersection of the profiles) is marked. The statistical errors for the tidal radii are most likely dominated by the mass incompleteness factor. Varying this factor by \SI{20}{\percent} results in changes of $r_t$ at about a \SI{5}{\percent} level, or about \SI{0.5}{pc}.

\begin{figure}
        \centering
        \resizebox{1.0\hsize}{!}{\includegraphics[]{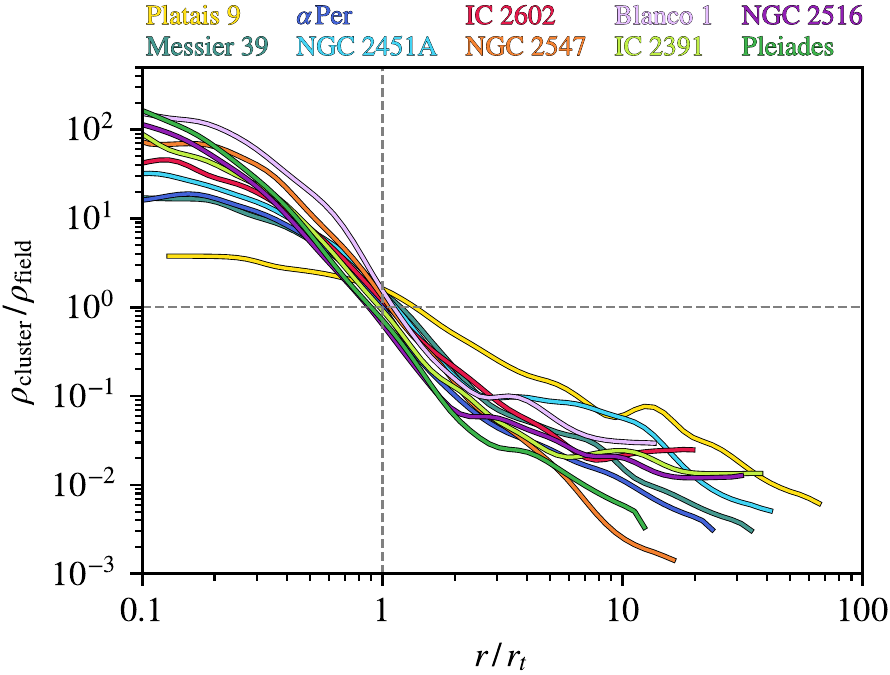}}
        \caption[]{Population volume densities normalized to the surrounding field as a function of distance from the cluster center, parameterized by the tidal radii. The observed volume density vanishes into the field at the tidal radius. Our member selection recovers sources across five orders of magnitude in volume density, down to three orders of magnitude below the Galactic field at the outermost rim of the coronae.}
    \label{img:volume_density}
\end{figure}

We also verified the determined tidal radii by comparing our selection to the surrounding Galactic field distribution. Specifically, the tidal radius should correspond to the radius, where the volume density of a population equals the density of the surrounding field. Similar to our approach to estimating the contamination fraction (Sect.~\ref{sec:methods:contamination}), we parameterized the volume density with the seventh nearest neighbor of each source. While for the contamination we only considered sources that survive the tangential velocity constraint, for this comparison we retained all \gaia sources in the vicinity of each group. Figure~\ref{img:volume_density} shows the ratio of the population volume density ($\rho_\mathrm{cluster}$) to the volume density in the field ($\rho_\mathrm{field}$) as a function of distance from the center of a group, normalized to the calculated value for the tidal radius. For all groups and within about 10\%, the determined tidal radii correspond to the radii where the cluster volume density drops below the density in the field. In addition, we observe that the volume density contrast of our selection compared to field stars ranges across about five orders of magnitude from well above 100 down to $10^{-3}$ for individual sources.

Inspecting Fig.~\ref{img:tidal_radii} more closely, we find subtle differences in the radial mass profiles. Several groups, and in particular, \aper, NGC~2547, Blanco~1, and IC~2391, show a pronounced kink in their profiles with very steep gradients toward the center. Other groups appear to have a much flatter profile with a less pronounced core structure and a smoother transition between the central regions and the coronae beyond the tidal radius. In reference to Fig.~\ref{img:volume_density}, the coronae themselves appear to follow a smooth distribution out to a few tidal radii before hitting the noise limit of our method with the current \gaia data release. As an exception to the general appearance of the profiles, even the central parts of Platais~9 appear to lie only marginally above mean field density, and the radial profile is significantly flatter as well.

\begin{figure}
        \centering
        \resizebox{1.0\hsize}{!}{\includegraphics[]{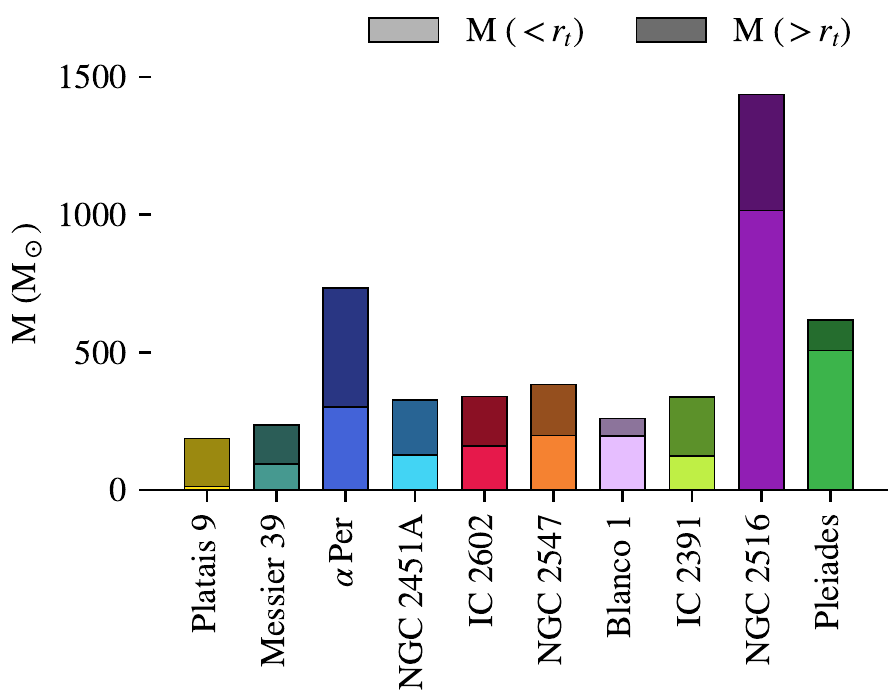}}
        \caption[]{Total mass content for each population. The statistics are displayed separately for the mass content inside and outside the tidal radius in light- and dark-shaded bars, respectively. For most clusters, the bulk of the mass content is located in their respective coronae.}
    \label{img:core_vs_tail_stats}
\end{figure}

Figure~\ref{img:tidal_radii} also shows that for several clusters, the enclosed mass fraction is well below \SI{50}{\percent} at the tidal radius. Comparing the measured tidal and half-mass radii in Table~\ref{tab:clusters_observables}, we find that the half-mass radii of only four populations (Blanco~1, NGC~2547, NGC~2516, and the Pleiades) are smaller than their tidal radii. This finding indicates that the bulk of the stellar mass content is located beyond the perimeter where the gravitational potential of the clusters dominates. The coronae are therefore largely gravitationally unbound with respect to the cluster cores. Figure~\ref{img:core_vs_tail_stats} visualizes the statistics on the mass fractions inside and outside the tidal radius, and the corresponding fractions are listed in Table~\ref{tab:clusters_observables}. Here, the statistics for NGC~2516 may be biased because we likely only recover parts of the trailing arm of its corona because of sensitivity limits and quality constraints on the data. Moreover, NGC~2547 has very similar half-mass and tidal radii, which leaves only Blanco~1 and the Pleiades with a clear excess of mass content inside their respective tidal radii (for Blanco~1, see also \citealp{Zhang20}).

% ============================================================================ %
\subsection{Cluster kinematics}
\label{sec:results:kinematics}

Our extensive membership determination also enables a more detailed look into the kinematic profile of each group. In particular, with the subset of sources that includes radial velocities in \gaia, we can determine 3D velocity vectors and subsequently characterize the expansion (or contraction) rate along different spatial axes. When the cluster samples are reduced to sources with all necessary measurements for calculating space motions, between about \SI{10}{\percent} and \SI{20}{\percent} of the full member list are retained, with a minimum of 30 sources in Messier~39 and a maximum of 242 sources for the Pleiades. Based on these subsets, we first constructed linear expansion profiles where we compared each spatial dimension ($XYZ$) with its corresponding velocities ($UVW$). A positive correlation between these components indicates expansion along the given spatial dimension, while a negative correlation indicates contracting motions.

\begin{figure}[!t]
        \centering
        \resizebox{1.0\hsize}{!}{\includegraphics[]{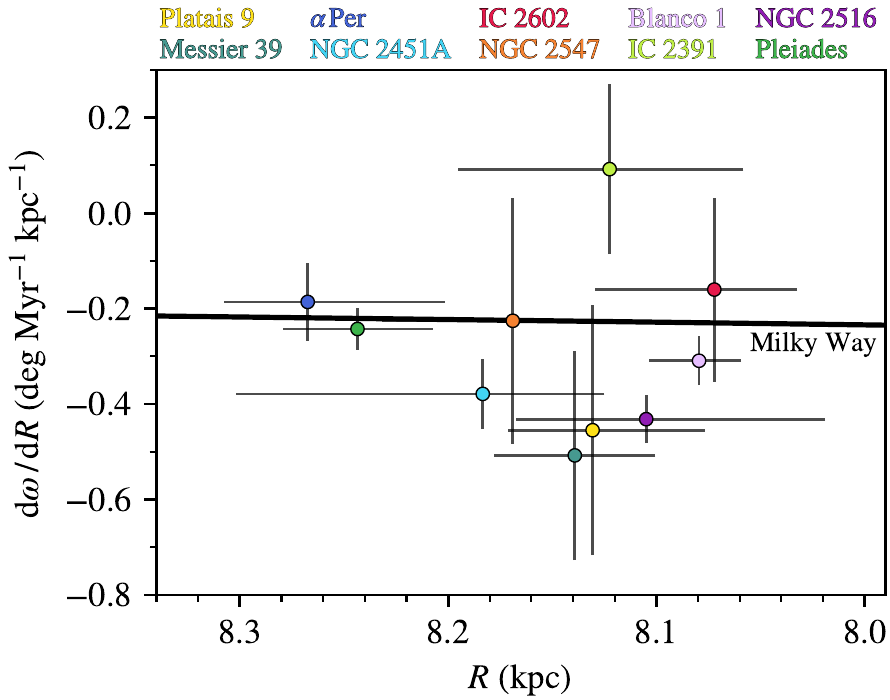}}
        \caption[]{Measured angular velocity gradients compared to the Milky Way rotation curve. Most populations show a significant gradient along their galactocentric radial axes that is largely compatible with or larger than the intrinsic velocity field of the Galaxy.}
    \label{img:differential_rotation_galaxy}
\end{figure}

The fitting procedure requires particular attention when linear relations between spatial and velocity components are constructed: the distribution of sources in the position-velocity planes (e.g., $X$-$U$) does not follow a linear correlation where the dispersion along the axes can be described with a normal distribution. Especially outliers with small measurement errors can affect a weighted fitting procedure, which leads to heavily biased slope determinations. Similarly, introducing another parameter to account for the scatter (as done by, e.g., \citealp{Wright19}) does not yield reliable results either because this additional scaling factor is different for all sources. For this reason, we decided to iteratively clip outliers prior to fitting the data by removing (a) sources with low signal-to-noise ratios (e.g., $|U/\sigma_U| < 3$) and (b) all sources outside a 3$\sigma$ range about the median in each velocity component. During the subsequent fitting, we then ignored measurement errors and determined the significance of the slopes by bootstrapping with 500 iterations. The fits themselves were obtained with the \textit{astropy} linear least-squares fitting engine.

\begin{figure*}[!t]
        \centering
        \resizebox{1.0\hsize}{!}{\includegraphics[]{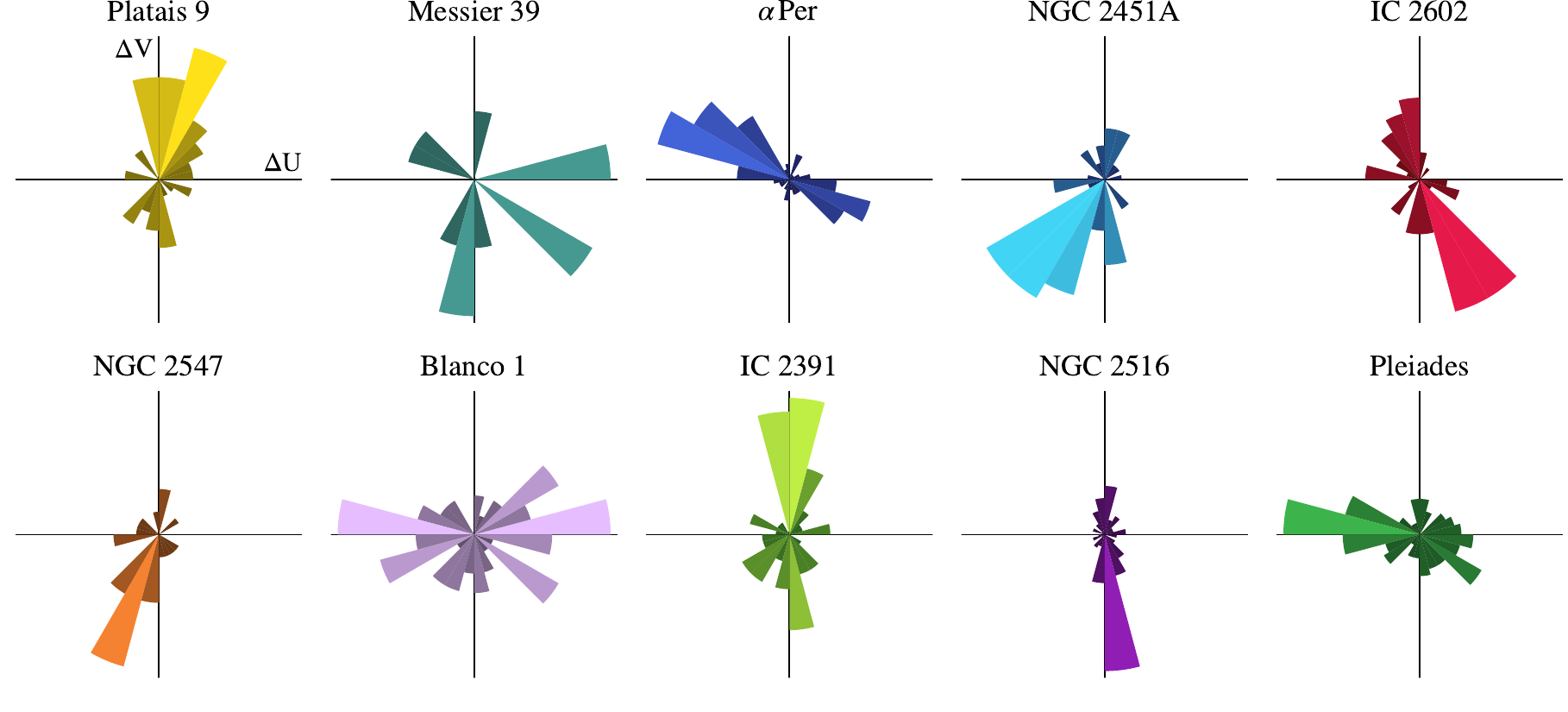}}
        \caption[]{$UV$ position angle histogram relative to the bulk motion of each cluster. Each wedge covers an angle of \SI{15}{\degree} , and its size and face color are proportional to the number of sources in the given direction (lighter colors and longer wedges denote more sources). Most clusters show a preferential symmetric axis of motion, but with an overall asymmetric internal velocity field.}
    \label{img:velocity_symmetry}
\end{figure*}

Table~\ref{tab:clusters_observables} lists the resulting slopes $k_{xu}$, $k_{yv}$, and $k_{zw}$ and their errors for the $X$-$U$, $Y$-$V$, and $Z$-$W$ planes, respectively. In addition, Fig.~\ref{img:linear_expansion} in the appendix shows the fits in all position-velocity combinations for each population, with the data values displayed relative to the group medians on both the abscissa and ordinate. The\ figure shows a clear correlation between the error magnitude and the dominating measurement parameter for a given velocity dimension. For instance, if the line of sight of a cluster aligns with the direction of motion in Galactic rotation, the $V$ component is mostly determined by radial velocity measurements. This is the case for Messier~39, where we find many $V$ measurements below our significance threshold. Similarly, Blanco~1 is located toward the direction of the Galactic south pole. As a consequence, the measured radial velocities mostly affect the vertical velocity $W$, which again leads to many statistically insignificant measurements in this component.

The determined slopes reveal several interesting findings. We first note that the statistical errors of the fits are relatively large. Because of the frequently large scatter in the position-velocity diagrams, this is an expected effect and emphasizes the reliability of our bootstrapped fitting errors. Nevertheless, all populations, except for NGC~2547 and the Pleiades, show slopes different from 0 in at least one dimension at a 2$\sigma$ significance level, six out of the ten clusters at a 3$\sigma$ level. Moreover, most statistically relevant slopes are positive, revealing largely expanding configurations. This is particularly the case for the $X$-$U$ plane, which means that all groups except for the Pleiades and to some degree also NGC~2547 appear to expand along this dimension. For Messier~39 and IC~2602, we furthermore observe negative slopes in the $Z$-$W$ plane, indicating contracting motions along the $Z$ -axis. For Messier~39, this measurement comes close to a 2$\sigma$ significance, while it exceeds the 3$\sigma$ level for IC~2602. This finding is of particular interest because both clusters show expanding motions along other spatial axes at the same time, revealing the highly dynamic states of the populations. Interestingly, we do not measure any significant expanding or contracting motions for the Pleiades: of our ten target populations, the fits for the Pleiades sources altogether result in values that are largely incompatible with expanding or contracting motions at all.

Because the cluster coronae stretch across hundreds of parsec, it is likely that the tidal field of the Galaxy, and specifically, the differential rotation pattern, affect the corona shapes and their dynamics. To investigate the nature of the expansion signature in reference to Galactic rotation, we further explored the relation between source kinematics and their position in the Galaxy. For this comparison, we chose to use angular velocities about the Galactic center (denoted as $\omega$) because of the large size of the cluster populations. In this particular case, angular velocities have a distinct advantage over space motions because sources with similar space velocities in the direction of Galactic rotation, but different galactocentric radii, will nonetheless drift apart over time as a result of different angular velocities.

For an effective comparison of our measurements, we computed the differential angular velocity gradient (d$\omega /$d$R$) of the clusters about the Galactic center and compared the results to the Milky Way rotation curve. These gradients were computed through linear fits of the angular velocities as a function of distance to the Galactic center (Fig.~\ref{img:differential_rotation}). Figure~\ref{img:differential_rotation_galaxy} displays the results of this procedure. We show the measured gradients in angular velocities for each cluster as a function of the Galactocentric radius. The indicated error in the ordinate is the statistical (1$\sigma$) error of the fit, while the error bar on the abscissa denotes the physical extent of each group. All populations, except for IC~2391, show an anticorrelation, in the sense of decreasing angular velocity with increasing Galactocentric radius. In the same figure, we show the gradient resulting from the differential rotation pattern of the Galaxy as derived from the rotation curve of the \code{MWPotential2014} potential \citep{galpy}. In this comparison, all clusters, except for IC~2391, clearly appear to have an angular velocity gradient significantly different from 0 that is either compatible with or larger than the intrinsic differential velocity field of the Galaxy. Interestingly, the clusters with values that exceed the differential rotation of the Galaxy itself are the oldest in our sample: Messier~39, Platais~9, and NGC~2516.

We furthermore explored the internal velocity distribution of each cluster to investigate the isotropy of the velocity field and to identify preferred directions of motion. Figure~\ref{img:velocity_symmetry} displays a position-angle histogram in the $UV$ plane relative to the bulk motion of the cluster core. Each wedge is a histogram bin with a size of \SI{15}{\degree}. All clusters, with the exception of Messier~39 because of the small number of sources, feature a well-visible preferential axis of motion. For instance, Blanco~1 and Pleiades members mainly move along the $X$ -axis, while stars in Platais~9 and NGC~2516 move mostly along the $Y$ -axis. At the same time, the velocity distribution frequently appears to be asymmetric. To test the symmetry of the velocity field, we performed a Wilcoxon signed-rank test \citep{Wilcoxon45} for each cluster and on each $UVW$ velocity component. For all clusters, except for Blanco~1 and IC~2391, we find statistically significant ($p < 0.05$) evidence for asymmetry in at least one velocity component. Furthermore, a comparison to Fig.~\ref{img:xyz} shows that there is no correlation between the dominant internal direction of motion and the spatial orientation of a cluster.

We also attempted to determine the isotropy in radial or tangential velocity components. Specifically, the anisotropy parameter $\beta$, such as used by \citet{Baumgardt07} or in a different form by \citet{Vesperini14}, is a popular parameterization in star cluster simulations to constrain formation and evolution physics. However, because of several biases in our measurements, we did not arrive at reliable values for $\beta$. Specifically, low number statistics, individual outliers, and large differences in measurement errors for individual velocity components resulted in values and errors for $\beta$ that did not allow any meaningful interpretation. For instance, using the definition of \citet{Baumgardt07}, our measured $\beta$ would have been compatible with both radial and tangential anisotropy within the statistical errors. On the other hand, the definition of $\beta$ by \citet{Vesperini14} includes measured velocity dispersions in each 3D velocity component. These measurements, however, are heavily affected by our perspective, where one component can be largely dominated by the measured heliocentric radial velocities, for instance, while the other components are mostly determined by the much better constrained proper motions. Conversely, determining velocity anisotropy from proper motions alone, as done by, for example, \citet{Wright19}, is not possible either because of the large on-sky extent of the coronae.

% ============================================================================ %
% ============================================================================ %
\section{Discussion}
\label{sec:discussion}

% ============================================================================ %
\subsection{Current cluster morphology}
\label{sec:discussion:morphology}

In Sect.~\ref{sec:results:spatial} we showed that the compact cluster cores are surrounded by large comoving and co-eval coronae (see Fig.~\ref{img:xyz}). To understand the observed morphology in general terms, we draw parallels to tidal structures of more evolved clusters and additionally take the observed kinematic profile into account. A particular characteristic pattern of the extended structures is their similar orientation in the $X$-$Y$ plane. The leading arm of all clusters with prominent coronae is always oriented toward the Galactic center. In reference to tidal structures in older star clusters, stars on outbound Galactic orbits form a trailing arm because the angular velocity decreases about the Galactic center. Conversely, stars on inbound galactic orbits overtake the cluster, thus forming a leading arm \citep{Kroupa08, ESSI}. Moreover, the measured angular velocity profiles agree well with this analogy, where we indeed find a matching gradient along the galactocentric radial axes (Figs.~\ref{img:differential_rotation_galaxy} and \ref{img:differential_rotation}). Consequently, stars that escape the gravitational potential of a cluster form highly elongated structures whose leading parts are tilted toward the Galactic center primarily as a consequence of Galactic differential rotation. Compared to this effect, the internal velocity field of each cluster can have a variety of preferred directions and appears to be negligible with respect to the shape of the formed coronae (Fig.~\ref{img:velocity_symmetry}).

Our results also illustrate that the characteristic tilt of the groups in the $X$-$Y$ plane is not universal: especially NGC~2451A appears to be distinct from the other populations. Because we find populations younger than NGC~2451A, we conclude that the larger tilt angle of this population is not a consequence of age. Furthermore, based on our measurements, we also conclude that NGC~2451A does not show unusual values regarding cluster kinematics, such as Galactic orbit, galactocentric radial velocity, or velocity dispersion. This means that differential Galactic rotation cannot be the only determining factor for the different observed tilts and elongations. The current shape of NGC~2451A may therefore be a consequence of birth conditions, including size and orientation of or velocity gradients within the birth molecular cloud, for instance. Moreover, the angular velocity gradients of some clusters are far higher than the value attributable to the rotation curve. As a consequence, some clusters have intrinsic velocity fields that amplify this signal so that their expansion rates exceed the inherent rate enforced by the Milky Way.

In general, we do not see a direct correlation between the age of a population and its current spatial arrangement: even in the youngest clusters, we observe a variety of shapes and sizes. For instance, IC~2391 and NGC~2451A already appear as heavily elongated structures. In contrast, the only marginally younger cluster NGC~2547, with an age of about \SI{30}{Myr} the youngest cluster in our sample, features a less pronounced corona. Taking this argument further, the \SI{300}{Myr} cluster Messier~39 is surrounded by a smaller coronal structure than some of its counterparts in the sample that are about five to ten times younger. This finding is not surprising because environmental and physical conditions at birth and during the earliest stages of the cluster evolution are important.

We also see significant overlap between the populations, where particularly IC~2602, IC~2391, and NGC~2451A form an intertwined network. Because we already observe such a significant overlap with only ten analyzed clusters and conservative selection criteria, we can only speculate about the complex 3D picture that all populations in this volume must portray. Our findings also offer a new perspective on our earlier results published in this paper series, where \citet{ESSIII} also reported evidence for temporal overlapping stellar populations. Because of the rigid assumption of impact parameters of only \SI{20}{pc} of the authors, the conclusion on the encounter timescale of about \SI{250}{Myr} may be severely overestimated, and cohabitation events may instead occur much more often, perhaps even most of the time.

% ============================================================================ %
\subsection{Origin of the coronae}
\label{sec:discussion:origin}

The current appearance of the cluster superstructures in both spatial and kinematic aspects appears to be largely affected by differential rotation in the Galaxy. The emergence of the coronae in the first place, however, can be attributed to different mechanisms. In one scenario, the stars that we observe today in the coronae were initially part of a single, much more compact cluster and only became unbound after formation. In this case, the stars in the coronae gradually left the cluster core regions to steadily contribute to the growth of the tidal structure. In another less well-studied scenario, the initial configuration could have resembled a more complex configuration, where star formation occurred in a larger hub and the currently observed cluster core was already decoupled from other star-forming parts in the birth molecular cloud. If the velocity gradients in such a hub were small enough, the entire structure could have stayed together for several dozen million years and eventually have formed the large coronae that we observe today.

In the scenario of initially compact clusters, important factors are the initial configuration, and according to theory, the phase of residual gas expulsion and subsequent violent relaxation. In this case, a testable prediction refers to the spatial configuration of the systems, as well as to the observed velocity structure. Specifically, if the observed coronae are indeed a consequence of the dynamical evolution of initially compact clusters, the (remnant) cluster cores would indeed be found preferentially centered on the superstructures, and the surrounding coronae would appear to be largely symmetric (ignoring perturbations since their formation). Similarly, the velocity field should appear largely symmetric about the cluster core, indicating that stars leave the cluster along a preferred spatial axis. While we do find that the cluster core is preferentially located at the center of the coronae (see Fig.~\ref{img:xyz}), we find mostly asymmetric velocity distributions about the cluster cores (Fig.~\ref{img:velocity_symmetry}). However, these measurements could be biased by our workflow. In our method, we searched for member stars that are kinematically compatible specifically with the cluster cores. As a consequence, we largely enforce symmetry in tangential velocity space, which is certainly mirrored in 3D velocities to some degree. Because of this potential bias with respect to symmetry, our measurements of clearly asymmetric velocity fields gain even more emphasis. At the same time, however, it is not clear if this bias could be systematically carried over to the spatial domain because the spatial domain on its own should not be affected by this problem as our application of DBSCAN in $XYZ$ coordinates has no prior information on cluster locations or their shapes.

Prominent examples of detailed numerical investigations in reference to the long-term evolution of star clusters can be found in \citet{Baumgardt07}, the paper series by Bekdaulet Shukirgaliyev \citep{Shukirgaliyev17, Shukirgaliyev18, Shukirgaliyev19}, and \citet{Dinnbier20}. Morphologically, the models in \citet{Shukirgaliyev18} and \citet{Dinnbier20} are indeed reminiscent of the coronae presented in this paper. However, depending on the initial configuration (e.g., SFE, the ratio of $r_h$ to $r_t$, or the dynamical state before gas expulsion), the models allow for a wide range of observed conditions at different cluster ages. As a consequence of our lack of knowledge about the initial conditions, as well as of the large variety of possible outcomes (and under the assumption that this scenario is indeed true), we are currently not in a position to present unambiguous evidence that this scenario did indeed occur.

A distinctive example can be made of the Pleiades, which we observe to be very compact; more than \SI{80}{\percent} of its mass lies within the tidal radius of the cluster. With our methods we are able to recover sources down to volume number densities of about \SI{e-3}{pc^{-3}} (see Fig.~\ref{img:volume_density_clusters} in the appendix). Based on this already excellent recovery threshold, the current observed cluster shape is roughly compatible with multiple setups as reported by \citet{Dinnbier20}. Depending on the gas-expulsion timescale and the compactness of the initial cluster, models with SFE$\;=1/3$, $2/3$, and even without any primordial gas (SFE$\;=1$) could resemble our determined cluster shape for the Pleiades. For this particular case, an optimization of our member identification to further improve the recovery rate seems to be necessary to draw meaningful conclusions. An improvement by another order of magnitude down to volume densities \SI{e-4}{pc^{-3}} could reveal more details and facilitate a better comparison to the \citet{Dinnbier20} models. To reach such limits, however, it will likely not be enough to improve the precision of the \gaia measurements. A significant boost in the recovery rate would be achieved by including additional features in the parameter space, such as a reliably measured third velocity component, or photometric information and population ages. In particular, 4MOST \citep{4most}, which will be installed at the VISTA telescope in the near future, will likely provide radial velocities for a significant fraction of cluster members.

In another formation scenario, the observed superstructures could be remnants of initially larger configurations. In such a setting, the observed cluster cores did not form in isolation, but instead were part of a larger complex with multiple simultaneous star-forming events. As a consequence, the observed structures could have formed in an already extended arrangement that is only amplified by the differential Galactic rotation. From an observational point of view, recent discoveries of young stellar structures on scales of hundreds of parsec provide an interesting reference point. For instance, \citet{Bouy15} explored large stream-like configurations of OB stars that seem to connect many very young clusters and OB associations in the solar neighborhood. Embedded in these so-called blue streams, \citet{Beccari20} identified a \SI{35}{Myr} old relic stellar filament that seems to bridge the gaps between several known and newly identified clusters (e.g., NGC~2547, NGC~2451B, Collinder~132, and Collinder~140). Because of the extent and age of the structure, the authors argued that it is most likely a remnant of a star-forming filament and not a consequence of tidal forces (for another example, see \citealp{Jerabkova19}). This observational evidence highlights that the observed coronae could be a consequence of the filamentary, highly substructured nature of star-forming regions. In reference to \citet{Beccari20}, we note that NGC~2547 is part of their identified structure (see also \citealp{Cantat-Gaudin19b,Cantat-Gaudin19a}). Our analysis of this cluster, however, does not produce unambiguous evidence that the cluster is embedded in a larger structure. NGC~2547 does appear kinematically well distinct from the larger complex that includes NGC~2451B, Collinder~132, and Collinder~140: Comparing the mean 3D velocity of NGC~2547 to NGC~2451B, we find similar $v_\phi$ and $v_Z$ components, but a difference in $v_R$ of about \SI{5}{\km \per \second}.

From a theoretical point of view, \citet{Dinnbier20} also considered the scenario of star formation in a larger complex, but in a largely idealized setting. Their results indicate that in such a case, the stars that formed distributed throughout the molecular cloud would indeed form an elongated structure with sizes up to several hundred parsec in diameter. However, the bulk of these stars would be located below our volume density recovery threshold. Moreover, in their experiment, they focused on the Pleiades, and the above-described conditions only apply to very specific circumstances and furthermore are only presented for a population age of \SI{125}{Myr}. At earlier stages in the cluster evolution, it is very likely that some imprint of the initial configuration is better preserved and may very well be visible within our constraints.

% ============================================================================ %
% ============================================================================ %
\section{Conclusions and summary}
\label{sec:conclusions}

We have analyzed ten nearby, prominent young open star clusters with \gaia DR2: \aper, Blanco~1, IC~2602, IC~2391, Messier~39, NGC~2451A, NGC~2516, NGC~2547, Platais~9, and the Pleiades (Fig.~\ref{img:dss}). We started our investigation by demonstrating the challenges of reliably identifying comoving stellar populations in proper motion space in large regions of the sky (Figs.~\ref{img:allsky_proper_motion} and \ref{img:vd_comparison}). Modern machine-learning tools offer outstanding and easily deployed possibilities for analyzing the wealth of data coming from \gaia. However, consistency checks and reliable contamination estimates must be included to allow for physically meaningful interpretations. We introduced a novel cluster member identification tool that bypasses the intricacies coming from projection effects in proper motion space (Fig.~\ref{img:method}). Following the member identification, we deconvolved the spatial distribution with a mixture of Gaussians to mitigate the effect of measurement errors in geometric distances (Fig.~\ref{img:method_xd}). This method will be made publicly available in a follow-up paper.

\vspace{\baselineskip}
\noindent The main results of this work can be summarized as follows:

\begin{enumerate}

\item We found that except for the Pleiades, all target clusters are surrounded by a vast stellar halo that we refer to as the corona. These coronae, together with the cluster cores, form extended cluster populations with sizes $\gtrsim\,$\SI{100}{pc} (Figs.~\ref{img:xyz} and \ref{img:wcs_matrix}). The coronae are comoving with the cluster cores and appear coeval in the observational HRDs. 

\item The spatial appearance of the coronae is reminiscent of tidal structures found in older clusters whose leading arm is oriented toward the inner Galaxy and that have an approximately symmetric trailing arm. This similarity is further confirmed by comparing the angular velocities of the cluster members about the Galactic center to the rotation curve of the Milky Way. All clusters, except for IC~2391, show angular velocity gradients across the coronae that are compatible with or exceed the intrinsic differential rotation pattern of the Galaxy (Fig.~\ref{img:differential_rotation_galaxy}).

\item The extracted cluster populations present a remarkably clean main sequence with stellar multiple sequences for most of the clusters and have 3D velocity dispersions typical of young populations (median \SI{1.4}{\km \per \second}; Figs.~\ref{img:contamination} and \ref{img:cmds}). We determine a contamination rate in the extracted populations of a few percent (median 3\%). 

\item When we compared our selection to the Galactic stellar field, our method recovered cluster members down to volume densities between two and three orders of magnitude below the background stellar field (Fig.~\ref{img:volume_density}). This corresponds to about \SI{e-3}{pc^{-3}} in volume number density (Fig.~\ref{img:volume_density_clusters}).

\item We constructed mass functions of all extended structures (core and corona) (Fig.~\ref{img:mass_functions}). The total determined system masses enabled a robust calculation of the tidal and half-mass radii of the clusters (Fig.~\ref{img:tidal_radii}). The radial cluster profiles reveal that the bulk of the stellar mass for many of the clusters is located beyond their tidal radius (Fig~\ref{img:core_vs_tail_stats}). 

\item A concise analysis of the velocity field of each cluster reveals a highly dynamic picture. All clusters, with the exception of NGC~2547 and the Pleiades, show statistically significant expansion along at least one spatial axis. For Messier~39 and IC~2602, we find evidence for simultaneous expansion along their Galactic $X$ (approximately galactocentric radial) axis and contraction along their Galactic vertical axis. The fits are presented in Fig.~\ref{img:linear_expansion}. 
\item The velocity field of the extended populations is aligned along a spatial axis that is unique to each cluster. The general direction of motion with respect to the cluster cores appears to be asymmetric for all clusters, except for Blanco~1 and IC~2391 (Fig.~\ref{img:velocity_symmetry}).

\end{enumerate}

The clusters we presented are relatively young and span ages across an order of magnitude from about 30 to \SI{300}{Myr}. We argue that it is most plausible that our sample of cluster cores and coronae is the outcome of a combination of different effects. The factors that contribute to their current appearance likely include (a) the complex morphology and kinematics of star-forming molecular clouds, (b) the ability of young stars to preserve their natal kinematic properties, (c) the consequences of residual gas expulsion and subsequent violent relaxation, and (d) differential rotation in the Galactic disk.

For the youngest populations (NGC~2451A, IC~2602, IC~2391, and NGC~2547), stars that formed throughout the entire parental molecular cloud complex would still comprise large parts of the coronae in the form of relic structures. For the more evolved clusters with ages of about or above \SI{100}{Myr} (Platais~9, Messier~39, \aper, Blanco~1, NGC~2516, and the Pleiades), this natal imprint could be beyond our current detection abilities, and the observed extended structures would be largely made up of stripped cluster stars. 

Finally, our application of a new method to ten prominent nearby open star clusters revealed an unexpected and exciting new perspective on their spatial and dynamical structure. Our findings expose more questions than answers and open several avenues for follow-up research to extend our knowledge on the evolution of stellar structures in the Galaxy.

% ============================================================================ %
% ============================================================================ %
\begin{acknowledgements}
This work has made use of data from the European Space Agency (ESA) mission \gaia (\url{https://www.cosmos.esa.int/gaia}), processed by the \gaia Data Processing and Analysis Consortium (DPAC, \url{https://www.cosmos.esa.int/web/gaia/dpac/consortium}). Funding for the DPAC has been provided by national institutions, in particular the institutions participating in the \gaia Multilateral Agreement.
This research made use of Astropy\footnote{\href{http://www.astropy.org}{http://www.astropy.org}}, a community-developed core Python package for Astronomy \citep{astropyI, astropyII}.
This research has made use of "Aladin sky atlas" developed at CDS, Strasbourg Observatory, France \citep{bonnarel00}.
We also acknowledge the various Python packages that were used in the data analysis of this work, including NumPy \citep{numpy}, SciPy \citep{scipy}, scikit-learn \citep{scikit-learn}, Matplotlib \citep{matplotlib}, and Plotly \citep{plotly}.
This research has made use of the SIMBAD database operated at CDS, Strasbourg, France \citep{simbad}. This research has made use of the VizieR catalogue access tool, CDS, Strasbourg, France (\citealp{vizier}; DOI: \href{http://dx.doi.org/10.26093/cds/vizier}{10.26093/cds/vizier}).
\end{acknowledgements}

% ============================================================================ %
% ============================================================================ %
\bibliography{references}
\clearpage

% ============================================================================ %
% ============================================================================ %
\begin{appendix}

% ============================================================================ %
% ============================================================================ %
\section{Supplementary plots and tables}
\label{app:suppl}

In this part of the appendix, we present a range of plots and tables that mainly provide supplementary information to the main article. Table~\ref{tab:sources} contains the first ten sources of our selection for each cluster (the full table is available online). All raw \gaia data can be obtained by cross-matching either the \gaia~DR2 source ID or the set of equatorial coordinates. In addition, we also list a set of variables specific to our analysis. These include the distance $d$ that was used for the member identification (raw), as well as the finally adopted distance after extreme deconvolution (xd). We also list the Galactic Cartesian velocities $UVW,$ and for each source, the contamination fraction $f_c$ and the name of the associated cluster. Table~\ref{tab:ages} lists a set of unique age determinations for the ten clusters discussed in this manuscript. This list is not complete, but should indicate the large variance with respect to age determinations for the populations. Our adopted age in this manuscript can be found in Table~\ref{tab:clusters}.

\begin{figure*}
        \centering
        \resizebox{1.0\hsize}{!}{\includegraphics[]{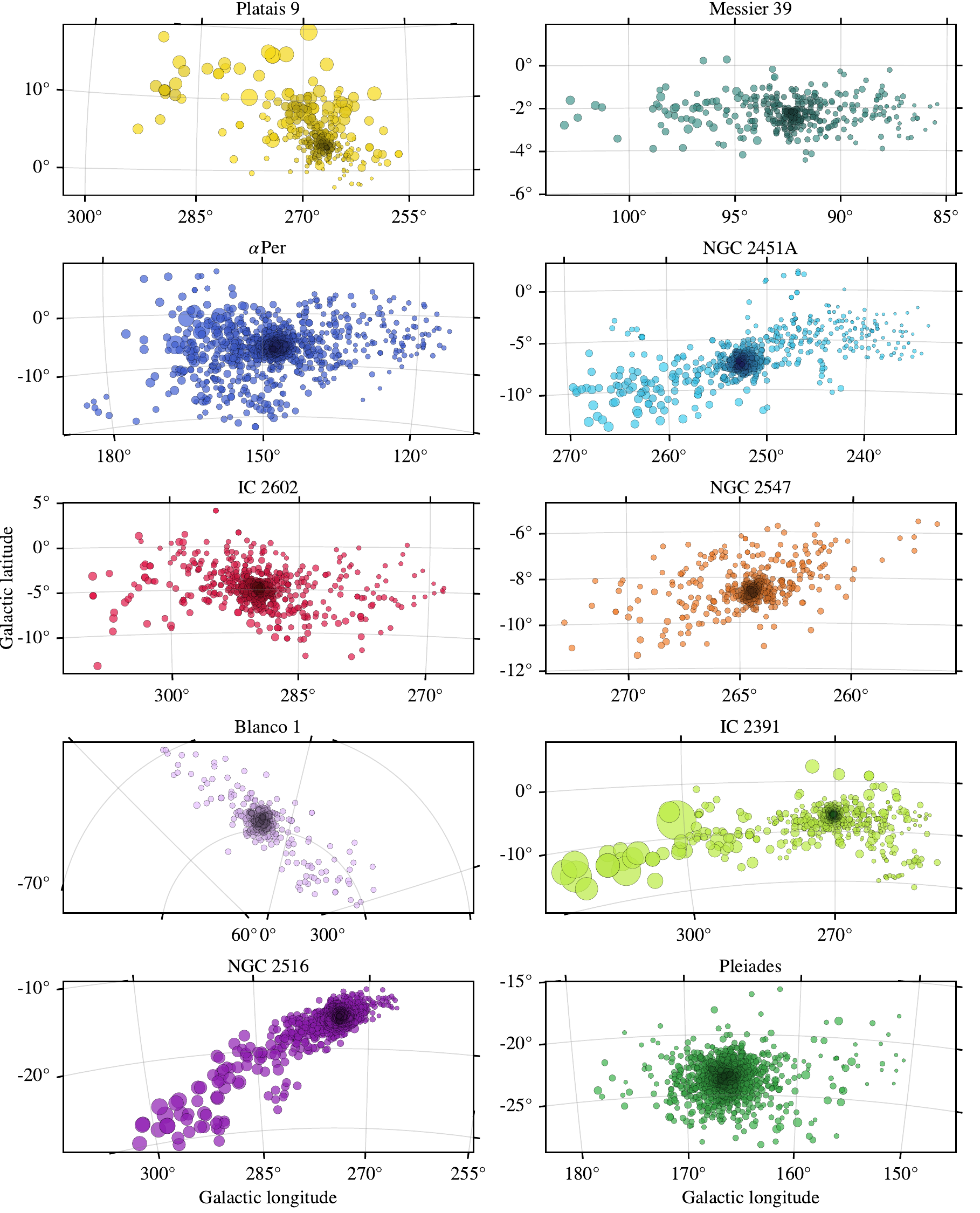}}
        \caption[]{On-sky view of clusters in Galactic coordinates in a stereographic projection. The luminance value of the symbol colors is proportional to volume density, with higher densities corresponding to darker shades. The symbol sizes are proportional to distance, where the size shrinks for increasing distances. The symbol sizes are scaled for each panel individually and are not comparable between clusters. }
    \label{img:wcs_matrix}
\end{figure*}

\begin{figure*}[!t]
        \centering
        \resizebox{1.0\hsize}{!}{\includegraphics[]{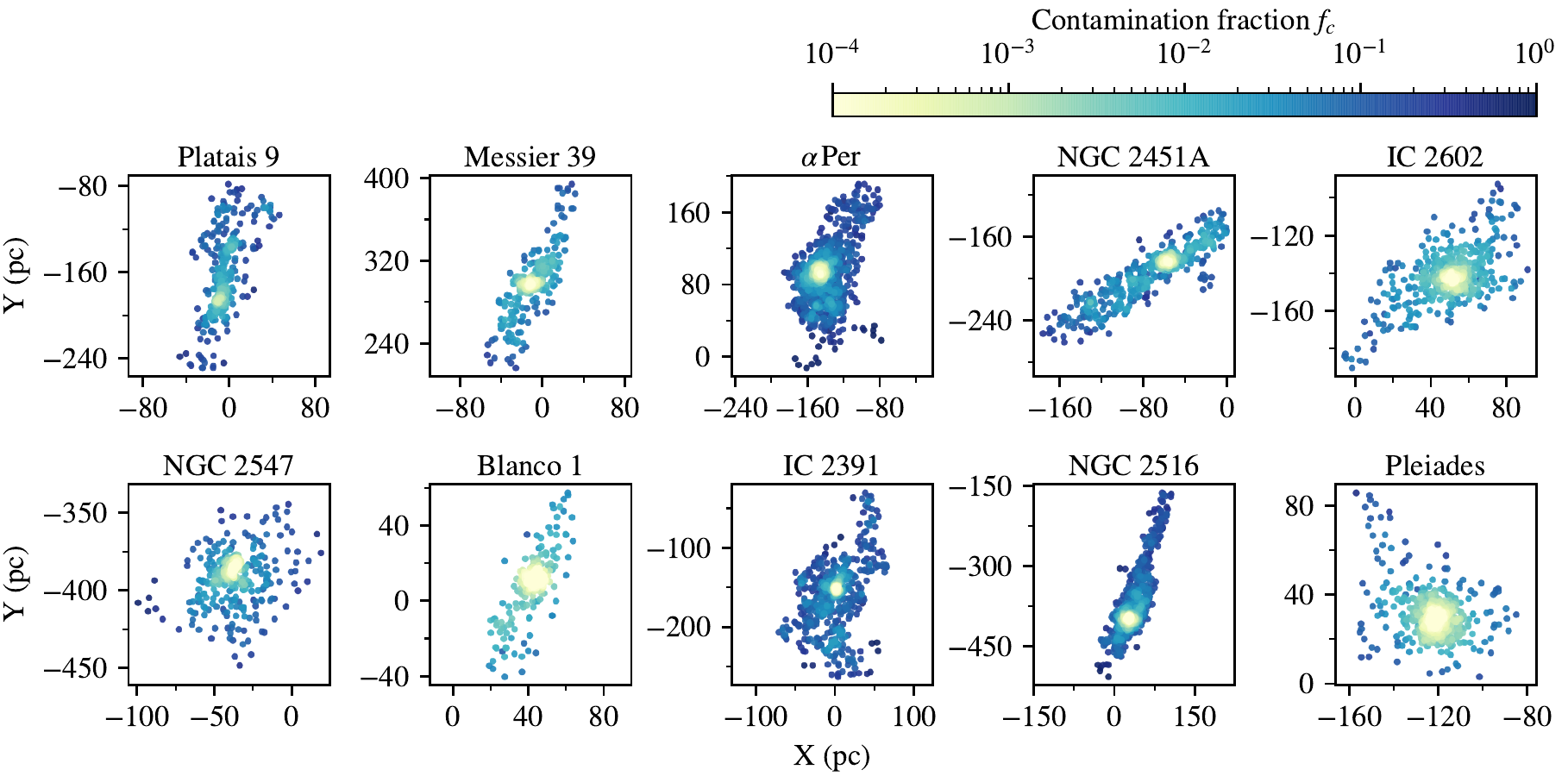}}
        \caption[]{Source distribution in the $XY$ plane (centered on the Sun) for each cluster, color-coded with the contamination fraction $f_c$. We observe a strong correlation between $f_c$ and the distance from the cluster centers, resulting from much lower volume number densities in the coronae.}
    \label{img:contamination_xy}
\end{figure*}

\begin{figure*}[!t]
        \centering
        \resizebox{1.0\hsize}{!}{\includegraphics[]{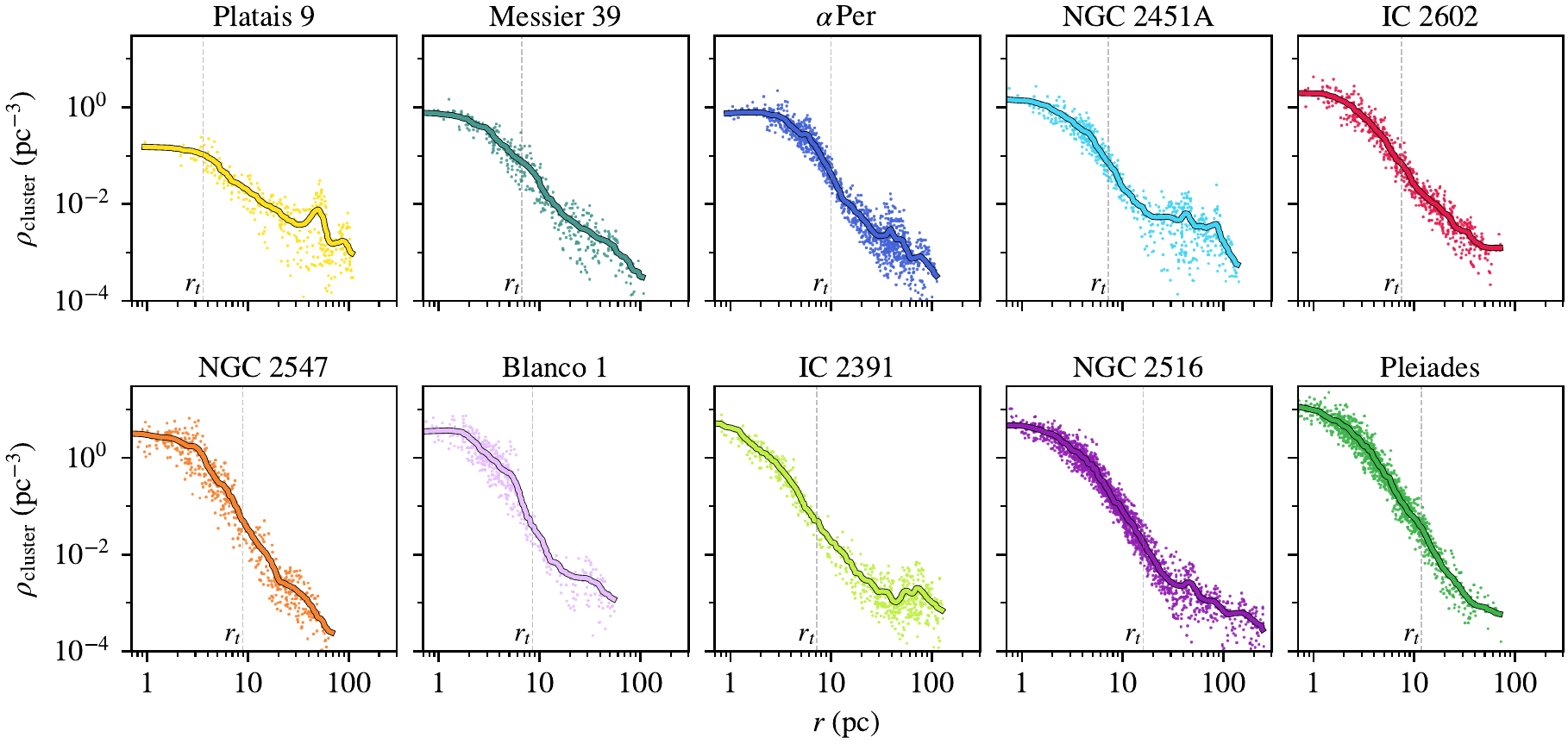}}
        \caption[]{Volume number density profiles as a function of radius from the density maxima of the clusters. The volume density was parameterized with the seventh nearest neighbor of each source. The solid colored lines represent a smoothed running median that is plotted in addition to the values for individual sources. For reference, the tidal radius of each cluster is marked with a vertical gray dashed line.}
    \label{img:volume_density_clusters}
\end{figure*}

\begin{figure*}
        \centering
        \resizebox{1.0\hsize}{!}{\includegraphics[]{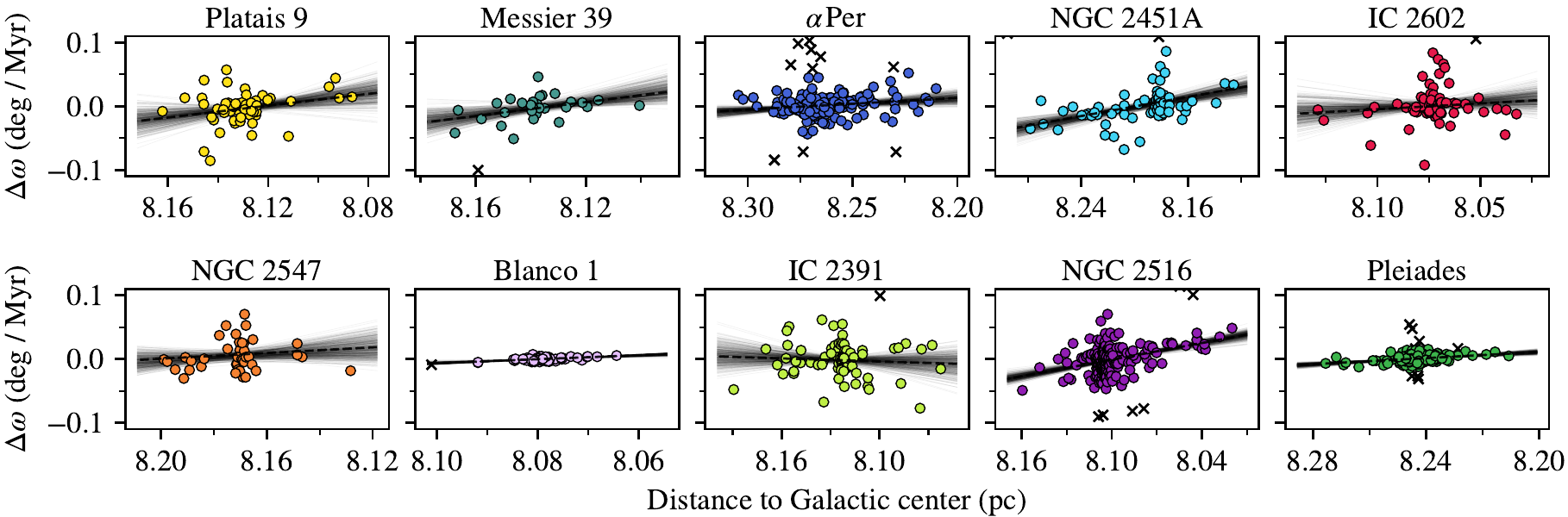}}
        \caption[]{Galactocentric angular velocity $\omega$ as a function of galactocentric radius for each cluster. The solid gray lines show 500 independent bootstrapped fitting results. The dashed lines are the adopted linear fits that were determined as the median in all realizations. The crosses are individual measurements below 3$\sigma$ significance and were not included in the fit.}
    \label{img:differential_rotation}
\end{figure*}

\begin{figure*}
        \centering
        \resizebox{1.0\hsize}{!}{\includegraphics[]{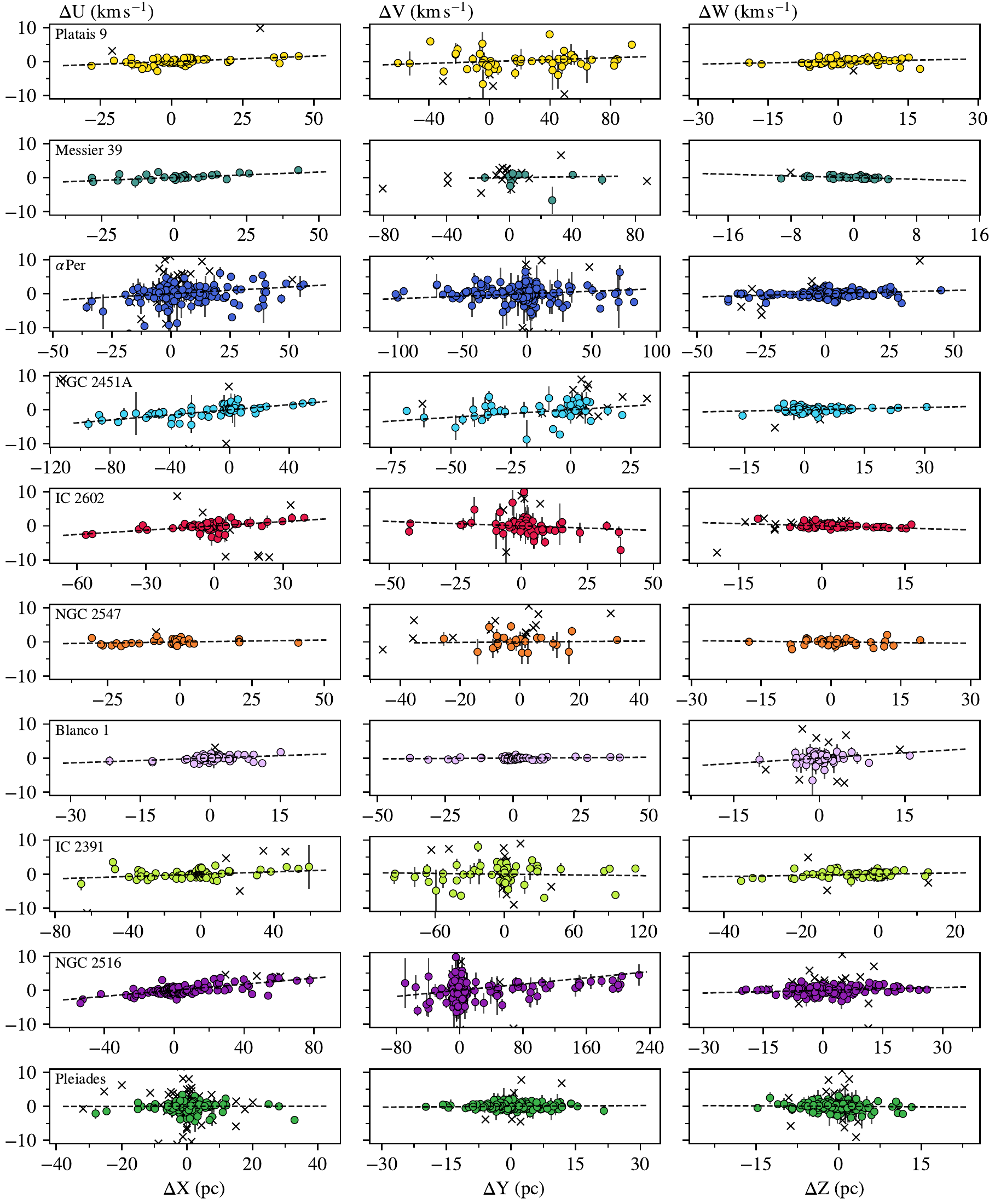}}
        \caption[]{Linear expansion model fits. Each row represents an individual cluster. The columns from left to right show the relation and linear fits for the parameter combinations $X$-$U$, $Y$-$V$, and $Z$-$W$, respectively. All values are shown relative to the density maxima or bulk motions of the clusters. The crosses are individual measurements below 3$\sigma$ significance and were not included in the fit.}
    \label{img:linear_expansion}
\end{figure*}

Figure~\ref{img:wcs_matrix} displays the selected sources in Galactic coordinates in a stereographic projection. Individual sources are displayed as colored circles. The sizes of the data points are proportional to the measured distance, where smaller circles represent more remote sources. This scaling is calculated individually for each panel and cannot be compared between the clusters. Figure~\ref{img:contamination_xy} displays for each cluster the source distribution in the Galactic Cartesian $XY$ plane. The color scale for each point shows the calculated contamination fraction for each source. This measure increases with distance from the density maxima of the clusters. Figure~\ref{img:volume_density_clusters} displays the measured volume number density as a function of distance to the cluster center. The measured volume density for each source was parameterized with the distance to the seventh nearest neighbor $d_{\mathrm{7NN}}$ so that $\rho_{\mathrm{cluster}} = 6 / (\pi d_{\mathrm{7NN}}^3)$ (including the source itself, there are eight sources in the given volume spanned by the distance to the seventh nearest neighbor). For all populations, we can recover sources down to about \SI{e-3}{pc^{-3}}. Figures~\ref{img:linear_expansion} and \ref{img:differential_rotation} provide the results of our linear fitting procedures for the $XYZ$-$UVW$ linear expansion model and angular velocity gradient about the Galactic center, respectively. In both diagrams, all sources included in the fit are displayed as colored points, whereas sources that did not pass the applied 3$\sigma$ significance threshold are marked as crosses.

\begin{table*}
\caption{First ten sources sorted by source ID. The distances raw and xd refer to the determined density maxima for the samples without and with applying the spatial deconvolution. $UVW$ are the velocity components in a Galactic Cartesian coordinate frame, and $f_c$ is the contamination fraction.}
\begin{tabular*}{\linewidth}{@{\extracolsep{\fill}} l c c c c c c c c c}
\hline\hline
Gaia DR2 source ID & RA & Dec & $d$ (raw) & $d$ (xd) & $U$ & $V$ & $W$ & $f_c$ & cluster \\
& (\si{\degree}) & (\si{\degree}) & (\si{pc}) & (\si{pc}) & (\si{\km \per \second}) & (\si{\km \per \second}) & (\si{\km \per \second}) &  &  \\
\hline
45022222315428608 & 56.11710 & 19.18482 & 145.03 & 144.90 & -2.65 & -28.88 & -11.82 & 0.016412 & Melotte\_22 \\
45612969296606080 & 63.39319 & 16.01258 & 136.74 & 136.62 & - & - & - & 0.121888 & Melotte\_22 \\
46342250447561216 & 61.44300 & 16.74528 & 122.76 & 123.91 & - & - & - & 0.053337 & Melotte\_22 \\
46760305383987456 & 62.20634 & 17.29149 & 153.11 & 153.01 & - & - & - & 0.118202 & Melotte\_22 \\
46843632046026240 & 61.97510 & 17.84030 & 123.73 & 123.77 & -5.86 & -28.68 & -12.12 & 0.044627 & Melotte\_22 \\
46922075328601216 & 62.80626 & 18.21945 & 169.49 & 169.35 & - & - & - & 0.167245 & Melotte\_22 \\
47508699141782656 & 65.27681 & 18.14764 & 127.66 & 127.86 & - & - & - & 0.105375 & Melotte\_22 \\
47573982644946688 & 63.88422 & 17.69776 & 126.10 & 126.49 & - & - & - & 0.070259 & Melotte\_22 \\
49488331171442560 & 64.11482 & 21.31705 & 159.61 & 158.52 & - & - & - & 0.098431 & Melotte\_22 \\
49809495943068288 & 59.45728 & 18.56219 & 129.97 & 129.97 & -6.80 & -28.94 & -14.02 & 0.010638 & Melotte\_22 \\
\hline
\end{tabular*}
\tablefoot{The full table is available at the CDS.}
\label{tab:sources}
\end{table*}

\begin{table*}
\caption{Compilation of cluster ages from the literature.}
\begin{tabular*}{\linewidth}{lcc}
\hline\hline
Cluster & ages & references \\
 & (Myr) &   \\
\hline
Platais 9 & 75-125, 79-347, 100, 123, 347 & 25, 26, 18, 13, 12\\
Messier 39 & 295-575, 371, 372, 891-1175 & 26, 13, 12, 3\\
\aper & 35, 35-71, 50, 52, 75-95, 90-130 & 12, 26, 13, 16, 1, 4 \\
NGC 2451A & 28-36/32-54, 40-80, 50-80, 58, 72-148 & 20, 19, 8, 12 and 13, 26\\
IC 2602 & 24-39/26-37, 30, 32, 41-52, 56-112, 68 & 20, 22, 24, 7, 26, 12 \\
NGC 2547 & 20-40, 26-45/26-49, 32-42, 34-36, 50, 78 & 9, 20, 17, 9, 12, 13\\
Blanco 1 & 91-562, 104-123, 209 & 26, 11, 12\\
IC 2391 & 26, 30, 35, 30-200, 30-49/36-51, 45-55, 48-58, 76 & 6, 22, 15, 26, 20, 1, 2, 12\\
NGC 2516 & 63, 70-84/77-85, 95-151, 100-150/100-220, 107, 120, 141, 150, 299 & 24, 20, 21, 14, 15, 12, 16, 10, 13 \\
Pleiades &  100, 100-170, 110-150, 115, 120, 125, 132-220, 141 & 16, 26, 1, 5, 12, 23, 27, 13\\
\hline
\end{tabular*}
\tablebib{
(1) \citet{Navascues_2004}; (2) \citet{Barrado_1999ApJ...522L..53B}; (3) \citet{cantat-gaudin18}; (4) \citet{Cummings_2018}; (5) \citet{Dahm_2015}; (6) \citet{De_Silva_2013}; (7) \citet{Dobbie_2010MNRAS.409.1002D}; (8) \citet{Huensch_2003}; (9) \citet{Jeffries_2005MNRAS.358...13J}; (10) \citet{Jeffries_1998MNRAS.300..550J}; (11) \citet{Juarez_2014}; (12) \citet{Kharchenko05}; (13) \citet{Kharchenko_2013yCat..35580053K}; (14) \citet{Lyra_2006}; (15) \citet{Mermilliod_1981A&A....97..235M}; (16) \citet{Meynet_1993A&AS...98..477M}; (17) \citet{Naylor_2006MNRAS.373.1251N}; (18) \citet{Platais_1998AJ....116.2423P}; (19) \citet{Platais_2001AJ....122.1486P}; (20) \citet{Randich_2018}; (21) \citet{Silaj_2014}; (22) \citet{Stauffer_1997ApJ...479..776S}; (23) \citet{Stauffer_1998ApJ...499L.199S}; (24) \citet{TADROSS_2002}; (25) \citet{Vande_2010}; (26) \citet{Yen_2018}; (27) \citet{Zhang_2019ApJ...887...84Z}}
\label{tab:ages}
\end{table*}

% ============================================================================ %
% ============================================================================ %
\section{Coordinate system definitions}
\label{app:coordinates}

The following list summarizes our parameter setup for the calculation of galactocentric cylindrical velocities:

\begin{itemize}
    \item $\alpha_{GC}$ = \SI{266.4051}{\degree}; $\delta_{GC}$ = \SI{-28.936175}{\degree}; right ascension and declination of the Galactic center \citep{galcen_coord}.
    \item $R_0$ = \SI{8122}{pc}; distance from the vantage point to the Galactic center \citep{galcen_distance}.
    \item $\phi_0$ = \SI{180}{\degree}; position angle of the Sun in galactocentric cylindrical coordinates.
    \item $Z_0$ = \SI{20.8}{pc}; height of the Sun above the Galactic plane toward the Galactic north pole \citep{z_sun}.
    \item $v_{R,\odot}$ = \SI{12.9}{\km \per \s}, $v_{\phi,\odot}$ = \SI{245.6}{\km \per \s}, $v_{Z,\odot}$ = \SI{7.78}{\km \per \s}; solar velocity in galactocentric cylindrical coordinates \citep{galcen_v_sun}.
    \item $(U, V, W)_{\mathrm{solar}}$ = (11.1, 12.24, 7.25) \si{\km \per \s}; barycentric velocity of the Sun relative to the local standard of rest \citep{vlsr}.
    \item $v_R$; galactocentric radial velocity component. Positive when moving away from the Galactic center.
    \item $v_{\phi}$; galactocentric azimuthal velocity component. Positive in the direction of Galactic rotation.
    \item $v_Z$; galactocentric vertical velocity component. Positive toward the Galactic north pole.
\end{itemize}

\end{appendix}

\end{document}